\documentclass[aps,prl,twocolumn,superscriptaddress,groupedaddress]{revtex4}  
\usepackage{graphicx,comment,float,braket,textcomp,amsmath,lipsum,color,soul}  
\usepackage{dcolumn}   
\usepackage{bm}        
\usepackage{amssymb}   

\begin{document}

\title{Multidimensional Terahertz Probes of Quantum Materials}

\author{Albert Liu}
\email{aliu1@bnl.gov} 
\affiliation{Condensed Matter Physics and Materials Science Division, Brookhaven National Laboratory, Upton, New York, 11973 USA}

\vskip 0.3cm

\date{\today}

\begin{abstract}
    Multidimensional spectroscopy has a long history originating from nuclear magnetic resonance, and has now found widespread application at infrared and optical frequencies as well. However, the energy scales of traditional multidimensional probes have been ill-suited for studying quantum materials. Recent technological advancements have now enabled extension of these multidimensional techniques to the terahertz frequency range, in which collective excitations of quantum materials are typically found. This Perspective introduces the technique of two-dimensional terahertz spectroscopy (2DTS) and the unique physics of quantum materials revealed by 2DTS spectra, accompanied by a selection of the rapidly expanding experimental and theoretical literature. While 2DTS has so far been primarily applied to quantum materials at equilibrium, we provide an outlook for its application towards understanding their dynamical non-equilibrium states and beyond.
\end{abstract}

\maketitle

The physics of quantum materials is one of the frontiers of modern condensed matter physics \cite{Keimer2017}. It presents a broad, formidable challenge, in particular due to the interaction of numerous degrees of freedom and their dynamical fluctuations, yet resolving these microscopic physics promises an unprecedented understanding of these quantum materials. How quantum materials cross phase boundaries in equilibrium, and how they are driven into new phases of matter with no equilibrium counterpart are questions essential to realizing their imaginable functionalities \cite{Tokura2017,Basov2017}. 

To this end, important messengers of the underlying physics are changes in their excitation spectra that accompany structural and electronic rearrangements, which typically occur in the terahertz ($10^{12}$ Hz) frequency range \cite{Nicoletti2016,Yang2023}. Yet our understanding and exploitation of material properties in the terahertz frequency range has been hindered by the challenges of generating, detecting, and manipulating light at these frequencies \cite{Davies2004} for spectroscopic applications. In particular, many aspects of coupling between different degrees of freedom, complex many-body interactions, and dynamical phenomena remain hidden in conventional terahertz spectroscopies sensitive only to two-point correlations of relevant observables.

In contrast, sophisticated techniques have been developed at microwave frequencies for nuclear magnetic resonance (NMR) spectroscopy \cite{Friebolin1991,Keeler2010} to access higher-order correlations of spin degrees of freedom. Various excitation schemes are routinely used to distill material nonlinearities, and have also been translated to the visible and near-infrared frequency ranges \cite{HammZanni2012,MDCS_Book}. In particular, multidimensional coherent spectroscopy at optical frequencies \cite{Cundiff2013} has emerged as the preeminent method for disentangling complex electronic dynamics in chemical and biological systems \cite{Fuller2015}, allowing spectroscopists to dissect material nonlinearities into their individual constituent terms \cite{Liu_2022}. 

Significant efforts have now been made to extend multidimensional techniques into the terahertz frequency range towards two-dimensional terahertz spectroscopy (2DTS) \cite{Lu2019}. Since its first demonstration over a decade ago \cite{Kuehn2009}, the growing accessibility of strong-field terahertz light sources has led to applications of 2DTS in a wide variety of material systems \cite{Lu2016,Maag2016,Houver2019,Mahmood2021,Pal2021,Lin2022,Luo2023,Blank2023,Liu_2023_echo,Zhang2024,Katsumi2024}. But perhaps their most intriguing applications lie in quantum materials, whose spectacular emergent properties arise precisely from the electronic correlations that 2DTS excels at resolving. In this Perspective, we introduce 2DTS as a powerful new technique for studying the intrinsic properties of quantum materials and discuss future opportunities for applying 2DTS as an ultrafast probe of quantum materials dynamically driven out of equilibrium with light \cite{Bao2022}.

\section{Basics of 2-D Terahertz Spectroscopy}

The basics of 2-D spectroscopy at infrared and optical frequencies are well-established and has been described extensively in the literature from a quantum level-system perspective \cite{HammZanni2012,MDCS_Book}. On the contrary, low-energy collective excitations of quantum materials are often described by effective coordinates and their equations of motion \cite{Nicoletti2016}. Our starting point will therefore be the generic potential energy of a coordinate $x$, which takes the form:
\begin{align}\label{GenericPotential}
    U(x) = \frac{1}{2}\omega_0^2x^2 + z^*xE(t) + U_{anh}
\end{align}
where $\omega_0$ is its resonance frequency dictated by the harmonic potential and $z^*$ is the effective charge that determines dipolar coupling of the coordinate to an external time-dependent electric field $E(t)$. In the following we provide a brief introduction to the nonlinearities of a classical oscillator and their corresponding 2-D spectra, which will then lend intuition to the unique physics underlying various features that we describe later. 

\begin{figure*}
    \centering
    \includegraphics[width=0.7\textwidth]{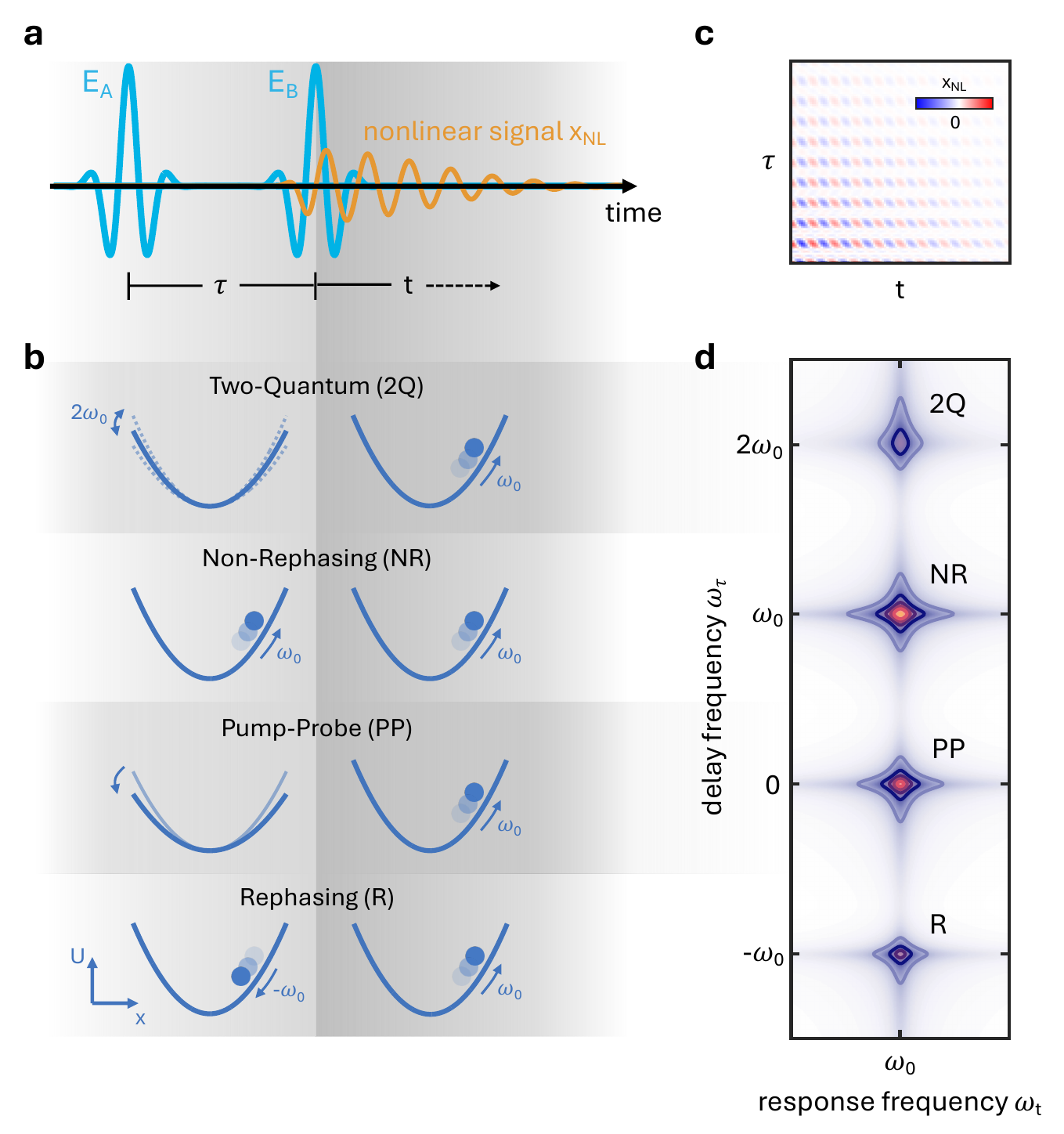}
    \caption{{\bf Classical view of two-dimensional terahertz spectroscopy.} {\bf a} Terahertz excitation pulse sequence involving two nominally identical fields $E_A$ and $E_B$ separated by a time delay $\tau$. A nonlinear signal $x_{NL}$ then evolves along the real laboratory time $t$. {\bf b} Illustrations of the four third-order nonlinearities with a response frequency of $\omega_0$ that result from a soft quartic anharmonicity. Note that the two remaining third-harmonic generation nonlinearities occur, but are not shown here. {\bf c} Nonlinear signal $x_{NL}$ in the time-domain as a function of $\{\tau,t\}$, which exhibits intricate oscillatory behavior due to interference between the multiple nonlinearities. {\bf d} Fourier transform of the time-domain signal in {\bf c}, showing four distinct peaks corresponding to the four nonlinearities illustrated in {\bf b} as indicated.}
    \label{Fig1}
\end{figure*}

\subsection{Self-Anharmonicity}

As a heuristic example, we take a quartic potential nonlinearity of the form:
\begin{align}\label{QuarticPotential}
    U_{anh}(x) = -\frac{1}{4}ax^4,
\end{align} 
where $a$ is an anharmonic coefficient that determines the strength of the nonlinearity. We will assume the case of a `soft spring' ($a > 0$), which describes, for example, the Josephson plasma resonance in layered superconductors \cite{Gabriele2022,Fiore2023,Salvador2024}.

In its most common realization, 2DTS involves two identical terahertz excitation pulses (denoted $E_A$ and $E_B$) that cooperatively drive a nonlinear response of the coordinate (shown in Figure~\ref{Fig1}a). This nonlinear signal is then generically a function of two variables, the inter-pulse time delay $\tau$ and the real laboratory time $t$ that evolves following interaction with the second pulse $E_B$. 
In the harmonic oscillator limit ($a = 0$), the superposition principle dictates that the coordinate response to $E_A + E_B$ is simply the summation of the individual responses to $E_A$ and $E_B$ alone. With the nonlinearity included however, the superposition principle is broken by a displacement-dependent resonance frequency \cite{Strogatz2024_Book}
\begin{align}\label{DuffingFrequency}
    \omega^2(x) = \omega_0^2 - ax^2.
\end{align}
Despite the simple nature of the underlying quartic potential, an infinite hierarchy of nonlinear coordinate responses scaling with $E_A^nE_B^m$ (where $n + m$ is odd) results to arbitrary order \cite{Liu2024_OptExpress}. Here we will only consider the most commonly measured third-order ($n + m = 3$) coordinate response composed of six contributing nonlinearities, four that emit at the fundamental resonance frequency $\omega_0$ and two that emit at triple its value $3\omega_0$ (third-harmonic generation). We further restrict our focus to only the four former terms, as the third-harmonic nonlinearities typically contain redundant information \cite{Salvador2024}.

To interpret such impulsive nonlinearities, it is most natural to consider the dynamics driven by $E_A$ and $E_B$ sequentially. Beginning with the arrival of $E_A$, its interaction with the system results in two primary consequences for the subsequent interaction with $E_B$ a time $\tau$ later (illustrated in Fig.~\ref{Fig1}b). First, it can immediately be seen from (\ref{DuffingFrequency}) that the resonance frequency is both rectified and modulated at twice the resonance frequency $2\omega_0$  \cite{Rajasekaran2016}. Second, $E_A$ introduces a time-dependent initial condition that either enhances or suppresses the response to $E_B$ (beyond a simple superposition of the two individual responses). To emphasize, these nonlinearities occur simultaneously and share a common response frequency $\omega_t = \omega_0$, resulting in overlapping signatures in one-dimensional spectra. As will be shown in the following however, this ambiguity is lifted in a two-dimensional spectrum.

\subsection{Two-Dimensional Spectrum}

To isolate the nonlinearities cooperatively driven by both $E_A$ and $E_B$ from their individual responses, one measures a nonlinear signal defined by
\begin{align}
    \nonumber x_{NL}(\tau,t) &= x(\tau,t)|_{Aon,Bon}\\ 
    &\hspace{0.7cm}- x(\tau,t)|_{Aon,Boff} - x(\tau,t)|_{Aoff,Bon}.
\end{align}
The equation of motion resulting from (\ref{QuarticPotential}) may be straightforwardly solved \cite{Liu2024_OptExpress} as a function of $\{\tau,t\}$ to obtain $x_{NL}(\tau,t)$, which exhibits complex oscillatory dynamics along both time axes as shown in Figure~\ref{Fig1}c. Interpretation of these dynamics is difficult in the time-domain, but becomes tractable in the frequency-domain. To this end, two-dimensional Fourier transform of $x_{NL}(\tau,t)$ correlates the dynamics along $\tau$ and $t$ in a two-dimensional (2-D) spectrum shown in Figure~\ref{Fig1}d. 

\begin{table}[b]
    \centering
    \begin{tabular}{ | c | c | c | }
        \hline
        {\bf Signal} & {\bf Dependence} & {\bf ($\omega_\tau$, $\omega_t$)} \\ \hline
        Two-Quantum (2Q) & $E_A^2E_B^*e^{i\omega_0(2\tau+t)}$ & $(2\omega_0,\omega_0)$ \\ \hline
        Non-Rephasing (NR) & $E_A|E_B|^2e^{i\omega_0(\tau+t)}$ & $(\omega_0,\omega_0)$ \\\hline
        Pump-Probe (PP) & $|E_A|^2E_Be^{i\omega_0t}$ & $(0,\omega_0)$ \\\hline
        Rephasing (R) & $E_A^*E_B^2e^{i\omega_0(t-\tau)}$ & (-$\omega_0,\omega_0)$ \\\hline 
        Third-Harmonic 1 (3H1) & $E_A^2E_Be^{i\omega_0(3t+2\tau)}$ & $(2\omega_0,3\omega_0)$ \\\hline
        Third-Harmonic 2 (3H2) & $E_AE_B^2e^{i\omega_0(3t+\tau)}$ & $(\omega_0,3\omega_0)$ \\\hline
    \end{tabular}
    \caption{Third-order wave-mixing nonlinearities}
    \label{table1}
\end{table}

In the two-dimensional (2-D) spectrum, four peaks appear at a response frequency $\omega_t = \omega_0$, corresponding to the four nonlinearities described above. Two peaks appear at delay frequencies $\omega_\tau = 0$ and $\omega_\tau = 2\omega_0$, corresponding to rectification and parametric modulation of the resonance frequency, and are denoted the `Pump-Probe' and `Two-Quantum' peaks respectively. The two other peaks appearing at $\omega_\tau = \omega_0$ and $\omega_\tau = -\omega_0$ then arise from in-phase or out-of-phase modulation (by the first excitation pulse $E_A$) of the nonlinear signal (stimulated by the second excitation pulse $E_B$), and are denoted the `non-rephasing' and `rephasing' peaks respectively. These nonlinearities, illustrated in Figure~\ref{Fig1}b and designated with terminology borrowed from atomic and molecular physics \cite{HammZanni2012,MDCS_Book}, are summarized in Table~\ref{table1} along with their excitation field scaling and frequency coordinates. We emphasize that, despite the apparent simplicity of (\ref{QuarticPotential}), these spectral features are generic for the third-order nonlinearity of an anharmonic oscillator and bear direct correspondence to their quantum level system counterparts. The classical description of their underlying mechanisms described here then lends intuition to how the various physics of quantum materials manifest in different features of a 2DTS spectrum, which will be the focus of the following section.

\section{2DTS of Quantum Materials: In Equilibrium}

In the previous section, we considered an anharmonic oscillator whose nonlinearity derives solely from a quartic potential nonlinearity. While instructive, this model can never realistically capture the physics of quantum materials. For example, collective excitations in real solids are dispersive, are affected by thermal fluctuations, couple to other degrees of freedom, and can suffer from material disorder. While such complications are difficult to disentangle with conventional spectroscopic probes, their effects naturally separate into unique features in a 2-D spectrum.

\subsection{Different Peaks, Different Physics}

\begin{figure}[t]
    \centering
    \includegraphics[width=0.49\textwidth]{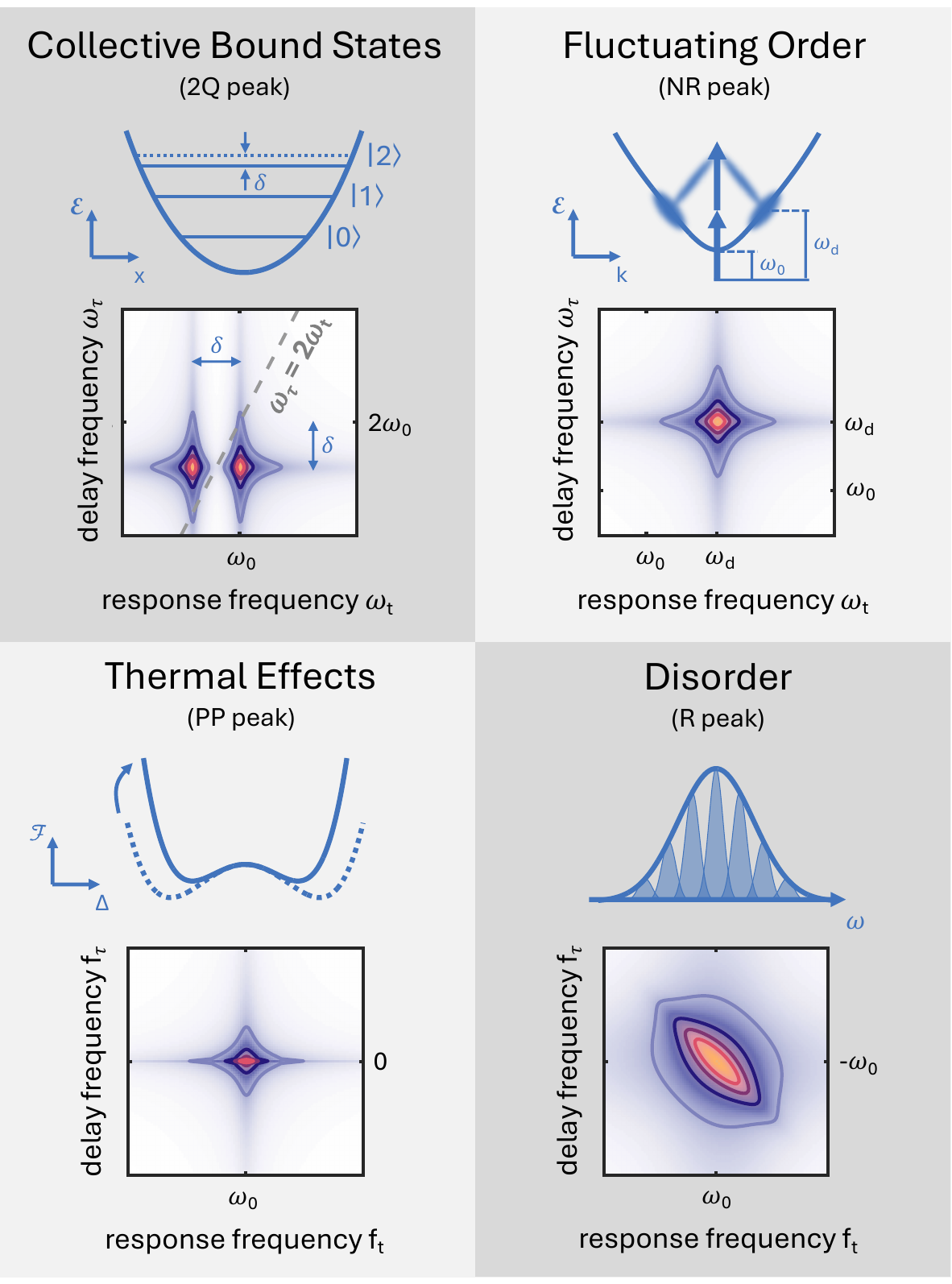}
    \caption{{\bf The unique physics of quantum materials revealed by each peak.} Top left: The two-quantum (2Q) nonlinearity is probes doubly-occupied states of a quantum oscillator, and exhibits unique signatures of quasiparticle bound states of binding energy $\delta$. Top right: The non-rephasing (NR) nonlinearity is sensitive (though not uniquely) to finite-momentum fluctuations, resulting in a peak position determined by the driving frequency $\omega_d$ rather than the resonance frequency $\omega_0$. Bottom left: The pump-probe (PP) nonlinearity captures all incoherent (non-oscillatory) processes. Most notably all thermal effects are isolated in this peak, leaving only non-thermal coherent dynamics elsewhere in the 2-D spectrum. Bottom right: The rephasing nonlinearity is unique in its sensitivity to disorder. In the absence of disorder the rephasing peak is symmetric while in the presence of disorder the rephasing peak elongates as shown.}
    \label{Fig2}
\end{figure}

In 2-D spectra of quantum materials, different aspects of their microscopic physics distort specific peaks away from their generic forms shown in Figure~\ref{Fig1}d. Examples of such phenomena are illustrated in Figure~\ref{Fig2} and summarized below for each peak.\\

{\it Two-Quantum (2Q)}: The two-quantum nonlinearity, as its name suggests, tracks dynamics evolving at (or near) twice the excitation photon frequency. These dynamics may arise classically, for example from parametric modulation of the resonance frequency (as illustrated in Fig.~\ref{Fig1}b). They may also be of quantum mechanical origin, arising from quantum coherences between two quasiparticle excitations and an unoccupied ground state \cite{Kim2009}. Such two-quantum nonlinearities have been observed, for example, for magnetic excitations in orthoferrites \cite{Lu2017,Huang2024} and Josephson plasmons in layered cuprates \cite{Liu_2023_echo}. In quantum materials, the two-quantum nonlinearity is particularly important for probing material anharmonicities \cite{Fulmer2004}, which can result in bound states of collective excitations such as bimagnons \cite{Wortis1963} or biphonons \cite{Cohen1969}. As illustrated in Fig.~\ref{Fig2}, for a binding energy $\hbar\delta$ exceeding the resonance linewidth, the two-quantum nonlinearity (1) shifts vertically along $\omega_\tau$ by $\delta$ and (2) splits horizontally along $\omega_t$ into two peaks separated by $\delta$. The two resultant peaks then correspond to a transition back into the unoccupied ground state ($\omega_t = \omega_0$) and dissociation of the collective bound state into single quasiparticles ($\omega_t = \omega_0 - \delta$).\\

{\it Non-Rephasing (NR)}: The non-rephasing nonlinearity is the strongest nonlinearity for a classical anharmonic oscillator, as in-phase excitation between $E_A$ and $E_B$ cooperatively drive the coordinate to the largest amplitude (strongest effect of anharmonicity). Besides its role in generating absorptive 2-D spectra (useful when conventional 2-D spectra are congested with overlapping peaks \cite{Khalil2003}), the non-rephasing nonlinearity is also sensitive to fluctuating order. In a recent theoretical study, Salvador et al. have predicted \cite{Salvador2024} non-rephasing 2-D spectra of the Josephson plasma resonance to contain signatures of superconducting fluctuations. To see this, one must consider a nonlinear excitation process (beyond the mean-field approximation) illustrated in the top right panel of Figure~\ref{Fig2}, in which two electric field interactions drive pairs of finite-momentum plasma waves at equal and opposite momenta \cite{Gabriele2021}. In the corresponding non-rephasing 2-D spectrum, a symmetric peak is observed whose peak position is fixed by the driving frequency $(\omega_\tau,\omega_t) = (\omega_d,\omega_d)$ due to energy conservation. This is in stark contrast to the spectrum predicted by mean-field theory for a damped oscillator, which exhibits a peak whose lineshape directly follows the linear loss function \cite{Salvador2024} (centered at the plasma frequency $\omega_0$). We anticipate the non-rephasing nonlinearity to also be sensitive to analogous processes of other types of excitations such as Klemens decay of phonons \cite{Klemens1966} or 3-magnon scattering \cite{Pirro2021}. \\

{\it Pump-Probe (PP)}: As evidenced by its scaling with $|E_A|^2$, independent of excitation phase, the pump-probe nonlinearity encompasses all incoherent (non-oscillatory) dynamics. Perhaps the most ubiquitous of such nonlinearities are thermal effects, in which initial excitation by $E_A$ (typically) raises the electronic temperature to a transient, non-equilibrium value. The bottom left panel of Figure~\ref{Fig2} illustrates such a thermal nonlinearity for an ordered system, in which the free energy is thermally-quenched by optical excitation. The corresponding pump-probe 2-D spectrum then exhibits a narrow lineshape along the vertical $\omega_\tau$ axis, reflecting the long timescales of thermal relaxation processes. A recent experiment by Kim et al. on the cuprate superconductor La$_{2-x}$Sr$_{x}$CuO$_4$ \cite{Kim2024} exploited this ability of 2DTS to separate coherent and incoherent dynamics, observing a strong pump-probe nonlinearity due to incoherent breaking of Coooper pairs \cite{Puviani2023} that is spectrally separated from the coherent order parameter dynamics. We note that in the case of dominant incoherent optical processes (thermal or otherwise) the pump-probe nonlinearity will appear alone in 2-D spectra, which has been observed in varied systems \cite{Folpini2017,Pal2021,Luo2023,Barbalas2023}. \\

{\it Rephasing (R)}: The rephasing nonlinearity is most well-known for its usefulness in studying disorder (being responsible for optical photon echoes \cite{Abella1966} and spin echoes in nuclear magnetic resonance \cite{Hahn1950}), which may be understood from either a time- or frequency-domain perspective for a disordered ensemble of oscillators (valid in the limit of strong localization). In the time-domain, the first pulse $E_A$ coherently excites dynamics of the ensemble which dephase due to disorder, obscuring their intrinsic lifetime. Arrival of the second pulse $E_B$ then performs a time-reversal operation, reversing the extrinsic dephasing and `rephasing' the ensemble after a time $t = \tau$. The bottom right panel of Figure~\ref{Fig2} then illustrates the frequency-domain perspective with both a 1-D spectrum and rephasing 2-D spectrum. In the 1-D spectrum, the resonance lineshape is broadened by a combination of both intrinsic and extrinsic broadening and one cannot disentangle the two effects. In the 2-D spectrum, these two broadening mechanisms are projected into orthogonal directions and produce an asymmetric `almond' peak shape \cite{Siemens2010,Liu2021_MQT,Salvador2024}. Recent 2DTS experiments have exploited this rephasing nonlinearity to measure intrinsic linewidths in doped silicon \cite{Mahmood2021} and to characterize disordered superconductivity in the cuprate La$_{2-x}$Sr$_{x}$CuO$_4$ \cite{Liu_2023_echo}. Analogous theoretical proposals have also been put forth toward resolving fractional excitations in spin liquids \cite{Wan2019,Li2021}.

\begin{figure*}
    \centering
    \includegraphics[width=1\textwidth]{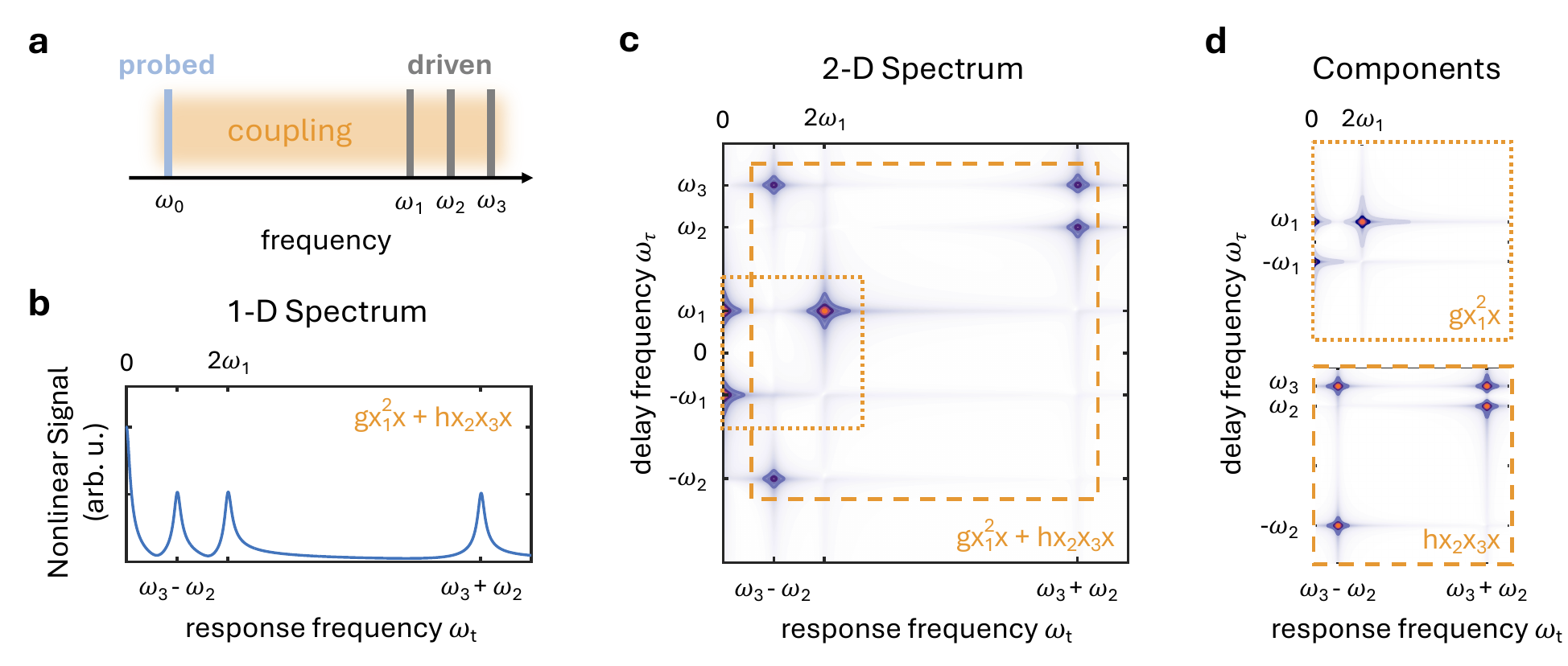}
    \caption{{\bf Dissecting anharmonic coupling using 2DTS.} {\bf a} Depiction of the probed coordinate $x$ with a resonance frequency $\omega_0$ indirectly excited through coupling to three driven modes $\{x_1,x_2,x_3\}$ with resonance frequencies $\{\omega_1,\omega_2,\omega_3\}$. Schematic {\bf b} one-dimensional and {\bf c} two-dimensional spectra of the nonlinear response resulting from the total anharmonic coupling $U_{anh} = gx_1^2x + hx_2x_3x$. The one-dimensional spectrum exhibits peaks at various response frequencies $\omega_t$ that are ambiguous to the underlying coupling mechanisms, while the 2-D spectrum provides more intricate peak patterns that inform the underlying drive frequencies. {\bf d} Decomposition of the total spectrum shown in {\bf c} into its two constituent components, which exhibit characteristic peak patterns used to identify the coupling term responsible. Note that the schematic spectra shown in {\bf b}, {\bf c}, and {\bf d} are obtained under a driven condition, in which the nonlinear driving force exceeds the oscillation period of $x$. In the inverse condition, impulsive excitation of $x$ becomes possible and additional features along $\omega_t = \omega_0$ appear.}
    \label{Fig3}
\end{figure*}

\subsection{Coupled Resonances}

At this point we have considered only a single driven oscillator and its self-anharmonicities \cite{vonHoegen2018}. In real quantum materials however, anharmonic coupling between different resonances plays an essential role in their material properties and is also responsible for many of their most intriguing behaviors \cite{Forst2015}. Such coupling is intricate, usually involving many degrees of freedom, and is difficult to decipher with one-dimensional spectroscopies. Disentangling the underlying coupling mechanisms in quantum materials is therefore a primary application of 2DTS, as we will demonstrate in the following.

As an example, we consider lowest-order anharmonic coupling (in a centro-symmetric system) between a coordinate $x$ and three coupled coordinates $\{x_1,x_2,x_3\}$, namely quadratic-linear \cite{Forst2011} and tri-linear \cite{Juraschek2017} coupling of the form:
\begin{align} \label{CouplingHamiltonian}
    U_{anh}(x_1,x_2,x_3,x) &= gx_1^2x + hx_2x_3x
\end{align}
where $g$ and $h$ are the anharmonic coefficients and the coupled coordinates $\{x_1,x_2,x_3\}$ have resonance frequencies $\{\omega_1,\omega_2,\omega_3\}$ respectively. Symmetry considerations then require the coordinates $\{x_1,x_2,x_3\}$ to be infrared-active and the commmon coordinate $x$ to be Raman-active ($z^* = 0$). As illustrated in Figure~\ref{Fig3}a, $\{x_1,x_2,x_3\}$ are directly driven by the excitation fields while the common coordinate $x$ is indirectly driven by anharmonic coupling and the corresponding force 
\begin{align} \label{CouplingForce}
    F_x(x_1,x_2,x_3) = -\frac{\partial U_{anh}}{\partial x} = -gx_1^2 - hx_2x_3,
\end{align}
where the former term gives rise to both a rectified ($\omega = 0$) and second-harmonic ($\omega = 2\omega_1$) response while the latter term results in sum/difference frequency ($\omega = \omega_3 \pm \omega_2$) responses. 

In the most general one-dimensional measurement of coupling in this system, an initial pulse resonantly excites all three coordinates $\{x_1,x_2,x_3\}$ before a subsequent pulse measures the response of $x$ via an observable such as transient reflectivity or transient birefringence. We further assume a driven condition, in which the nonlinear driving force exceeds the oscillation period of $x$. As shown in Figure~\ref{Fig3}b, the resultant one-dimensional measurement yields, however, only the response frequencies while providing minimal information on the underlying coupling mechanisms.

In contrast, the corresponding 2-D spectrum shown in Figure~\ref{Fig3}c is far richer. In addition to the response frequency axis, the dynamics are now spectrally resolved along the delay frequency axis $\omega_\tau$ to produce two-dimensional peak structures characteristic of specific coupling mechanisms. In Figure~\ref{Fig3}c, two sets of peaks (outlined by the dotted and dashed boxes) can be identified as corresponding to the two coupling terms in (\ref{CouplingHamiltonian}). The peak pattern of each component, plotted individually in Figure~\ref{Fig3}d, may then be intuitively understood by considering their coordinates along the two frequency axes $\omega_\tau$ and $\omega_t$.

Given a set of peaks at a particular response frequency $\omega_t$, their corresponding coordinates along the delay frequency $\omega_\tau$ indicate the parent driving frequencies that derive from (\ref{CouplingForce}). This may be most easily understood in the spectrum of $U_{anh} = hx_2x_3x$ (bottom panel of Figure~\ref{Fig3}d), where two peaks at $\omega_t = \omega_3 + \omega_2$ appear at corresponding delay frequencies $\omega_\tau = +\omega_3$ and $\omega_\tau = +\omega_2$ that unambiguously indicate a sum-frequency generation process. The two peaks at $\omega_t = \omega_3 - \omega_2$ likewise appear at $\omega_\tau = \omega_3$ and $\omega_\tau = -\omega_2$ to indicate a difference-frequency generation process. These fingerprints of sum- and difference-frequency generation processes were recently observed, for example, between magnon modes in YFeO$_3$ \cite{Zhang2024_2}.

The spectrum of $U_{anh} = gx_1^2x$ (top panel of Figure~\ref{Fig3}d) (which can be intuited as the $U_{anh} = hx_2x_3x$ spectrum in the degenerate limit $\omega_3 = \omega_2$), is interpreted in an analogous fashion. Here, two peaks at $\omega_t = 0$ appear at oppositely-signed frequencies $\omega_\tau = +\omega_1$ and $\omega_\tau = -\omega_1$ to indicate a rectified response while a single peak appears at $\omega_t = 2\omega_1$ and $\omega_\tau = +\omega_1$ due to second-harmonic generation of a single frequency.

The peak patterns of the two lowest-order coupling terms considered here may be straightforwardly generalized to other possible coupling mechanisms in quantum materials. For example, recent experiments have demonstrated 2DTS as an incisive probe of `activated' coupling involving phonons \cite{Blank2023}, magnons \cite{Zhang2024}, and even between magnons and phonons \cite{Mashkovich2021}. Higher-order coupling terms \cite{Fechner2024} should provide even more distinct signatures in 2-D spectra while other more exotic coupling mechanisms (not expected from straightforward symmetry considerations \cite{Khalsa2021}) will also be interesting directions of future studies.

\section{2DTS of Quantum Materials: Out of Equilibrium}

So far we have considered quantum materials in thermodynamic equilibrium, in which a multitude of quantum phases can be realized by varying {\it static} parameters such as temperature, pressure, and composition. In contrast to these traditional methods, recent years have witnessed rapid progress in {\it dynamic} manipulation of quantum materials with light \cite{Disa2021}. Light-induced ferroelectricity \cite{Nova2019,Li2019}, magnetism \cite{Wang2022,Disa2023}, topological switching \cite{Sie2019,Vaswani2020,Luo2021}, and even putative superconductivity \cite{Fausti2011,Mitrano2016,Fava2024} are all examples of non-thermal phases that can emerge when quantum materials are pushed out of equilibrium \cite{delaTorre2021}. Below, we provide an outlook for applying 2DTS toward understanding these light-induced phases, as well as the complex interplay of numerous degrees of freedom that underlie many of these spectacular effects.

\begin{figure*}
    \centering
    \includegraphics[width=0.99\textwidth]{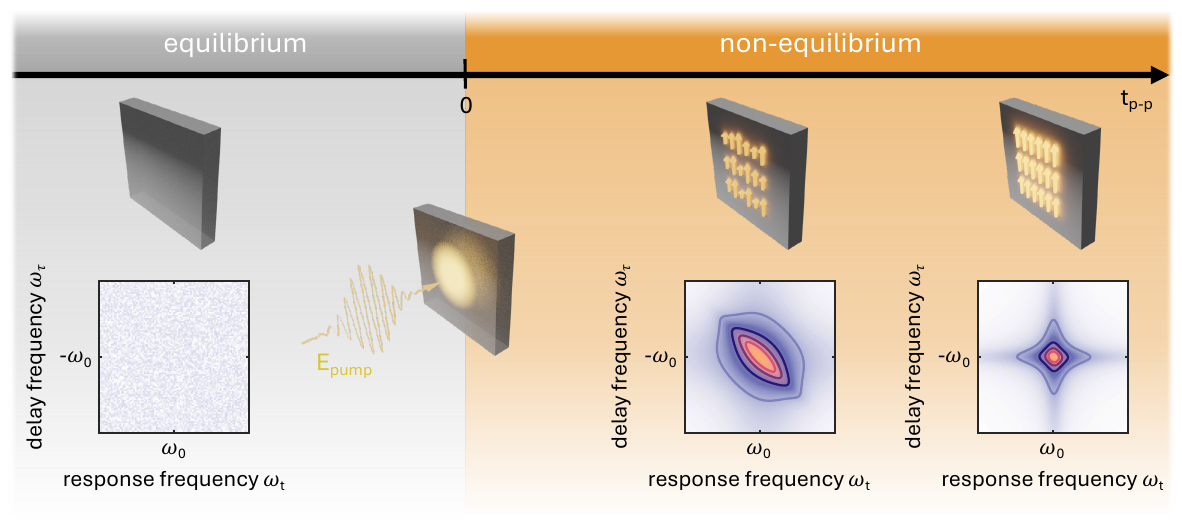}
    \caption{{\bf Pump-2DTS probe spectroscopy of a hypothetical non-equilibrium phase transition.} Evolution of the nonlinear optical response is shown as a function of pump-2DTS probe delay $t_{p-p}$. Before arrival of the pump pulse $(t_{p-p} < 0)$, the system is unordered at thermal equilibrium and no rephasing nonlinearity is observed at the salient collective mode frequency $\omega_0$. Upon arrival of the pump pulse $E_{pump}$ $(t_{p-p} = 0)$, a non-equilibrium phase transition is initiated. At short timescales ($t_{p-p} > 0$) the non-equilibrium phase is disordered, reflected by an elongated peak in the rephasing 2-D spectrum. At long timescales ($t_{p-p} \gg 0$), the disorder resolves and the rephasing 2-D spectrum becomes correspondingly symmetric.}
    \label{Fig4}
\end{figure*}

\subsection{Pump-2DTS Probe Experiments}

In most demonstrations of light-induced phenomena, an initial `pump' pulse drives a system into a non-equilibrium state, which is then interrogated by a subsequent `probe' pulse that measures a one-dimensional (either linear or nonlinear) observable. In this language, one may generalize 2DTS as a multidimensional probe of transient states that measures its multidimensional nonlinear optical response.

We illustrate this idea for a hypothetical light-induced phase transition induced by a pump pulse $E_{pump}$, in which the final non-equilibrium phase percolates from an initial disordered phase at short timescales. Such ultrafast disorder can then be probed by the rephasing nonlinearity generated by a subsequent 2DTS pulse sequence resonant with a salient collective mode frequency $\omega_0$, illustrated in Fig.~\ref{Fig4} for increasing values of the pump-2DTS probe time-delay $t_{p-p}$. Prior to arrival of the pump pulse ($t_{p-p} < 0$), no ordered phase exists at thermal equilibrium and a rephasing nonlinearity is absent for the collective mode that heralds a non-equilibrium ordered phase. Immediately after excitation by the pump pulse ($t_{p-p} > 0$), a non-equilibrium phase forms whose disorder (and that of its concomitant collective mode) is reflected by an elongated rephasing 2-D spectrum \cite{Siemens2010,Salvador2024}. With further increasing $t_{p-p}$, the non-equilibrium phase becomes less disordered and the rephasing 2-D spectrum becomes correspondingly symmetric.

Besides the rephasing nonlinearity, the other nonlinearities measured by 2DTS can provide even more information on non-equilibrium phenomena. For example, the pump-probe nonlinearity can isolate thermal effects such as the melting of competing orders \cite{Fausti2011,Kogar2020} while the two-quantum nonlinearity (and its higher-order counterparts) can inform the build-up of non-thermal correlations \cite{Turner2010} or loss thereof. Taken together, the information revealed by a 2DTS probe should ultimately address fundamental questions about non-equilibrium phase transitions such as their dynamic universality classes and scaling \cite{Halperin2019}, or whether these equilibrium concepts even apply at all to such radically different timescales.

\subsection{Perturbative Pump Experiments}

In our final generalization of 2DTS in a non-equilibrium context, we incorporate the `pump' into part of the 2DTS excitation sequence. In contrast to the typical intense pump field that induces drastic changes in material properties (an overtly non-perturbative response), we consider relatively weak pump fields and an induced response that is perturbative. In this regime the microscopic mechanisms leading to light-induced non-equilibrium phenomena, more specifically the various coupling pathways between a driven mode and other material degrees of freedom, can be characterized.

A recent study reported by Taherian et al. \cite{Taherian2024} demonstrates the power of this methodology towards studying the underlying mechanism of the light-induced state of underdoped YBa$_2$Cu$_3$O$_{6+x}$, in which the appearance of a Josephson plasma edge upon optical excitation was attributed to formation of a non-equilibrium superconducting state. In these experiments, $E_A$ and $E_B$ were tuned to the closely-spaced apical oxygen phonon modes at 17 THz and 20 THz while the nonlinear observable was transient second-harmonic generation induced by the excitation fields \cite{vonHoegen2022}. The resultant 2-D spectra revealed unambiguous signatures of Josephson plasma currents cooperatively excited by both driven phonon modes, identified by the characteristic peak pattern of a difference-frequency generation process shown in Fig.~\ref{Fig3}.

Many other light-induced phenomena result from intricate underlying mechanisms ambiguous to conventional one-dimensional probes. Non-equilibrium superconducting-like states in organic solids \cite{Mitrano2016,Buzzi2020}, ferroelectricity in SrTiO$_3$ \cite{Nova2019,Lu2019}, and magnetism in YTiO$_3$ \cite{Disa2023} are all problems that may require a multidimensional probe to disentangle. The additional frequency dimension in 2-D spectra also provides a more efficient alternative for characterizing pump frequency resonances, typically performed in the frequency-domain by sweeping the excitation frequency \cite{Liu2020_PRX,Rowe2023}.

\section{Outlook}

In this Perspective, we introduced the emerging technique of two-dimensional terahertz spectroscopy as an incisive probe of quantum materials, both in and out of equilibrium. Future directions are numerous, but we conclude with three potential developments of particular interest to us.

(1) The coherent nature of 2DTS resolves both the real (dispersive) and imaginary (dissipative) components of the nonlinear optical response \cite{Li2006}, information which is typically discarded in most experiments thus far that analyze amplitude 2-D spectra. Developing the experimental and theoretical framework for analyzing complex 2-D spectra should connect nonlinear observables with their more familiar linear counterparts (optical conductivity, refractive index, etc.), and may even provide insight into the phase of an underlying many-body wavefunction in topological materials \cite{Ma2021}. 

(2) Extending scanning probes \cite{Cocker2021} to perform 2DTS with nanometer resolution and below is another frontier that combines the advantages of multidimensional spectroscopy with imaging capability \cite{Chen2019}. Directly correlating the spectral disorder measured by 2DTS with underlying spatial disorder will strengthen the crucial connection between material inhomogeneities and the resultant optoelectronic properties \cite{Vojta2019}. 

(3) Finally, 2DTS may also be implemented with ultrafast circuitry \cite{Gallagher2019,Wang2023} to reach frequencies below what is possible with free-space optical methods \cite{Yoshioka2024}. Such on-chip methods also circumvent the restrictive diffraction limit of terahertz light, and integrate naturally with studies of van der Waals materials and their devices \cite{McIver2020,Zhao2023}.

\section{Acknowledgments}

The author's perspective on multidimensional spectroscopy of quantum materials has been shaped by fruitful collaborations with many colleagues: Andrea Cavalleri, Eugene Demler, Alex Gomez Salvador, Pavel Dolgirev, Marios Michael, Michael Fechner, Michael F\"{o}rst, Niloofar Taherian, Alex von Hoegen, Ankit Disa, and Danica Pavicevic. Albert Liu was supported by the U.S. Department of Energy, Office of Basic Energy Sciences, under Contract No. DE-SC0012704.


\begin{thebibliography}{91}
\expandafter\ifx\csname natexlab\endcsname\relax\def\natexlab#1{#1}\fi
\expandafter\ifx\csname bibnamefont\endcsname\relax
  \def\bibnamefont#1{#1}\fi
\expandafter\ifx\csname bibfnamefont\endcsname\relax
  \def\bibfnamefont#1{#1}\fi
\expandafter\ifx\csname citenamefont\endcsname\relax
  \def\citenamefont#1{#1}\fi
\expandafter\ifx\csname url\endcsname\relax
  \def\url#1{\texttt{#1}}\fi
\expandafter\ifx\csname urlprefix\endcsname\relax\def\urlprefix{URL }\fi
\providecommand{\bibinfo}[2]{#2}
\providecommand{\eprint}[2][]{\url{#2}}

\bibitem[{\citenamefont{Keimer and Moore}(2017)}]{Keimer2017}
\bibinfo{author}{\bibfnamefont{B.}~\bibnamefont{Keimer}} \bibnamefont{and} \bibinfo{author}{\bibfnamefont{J.~E.} \bibnamefont{Moore}}, \bibinfo{journal}{Nature Physics} \textbf{\bibinfo{volume}{13}}, \bibinfo{pages}{1045} (\bibinfo{year}{2017}), ISSN \bibinfo{issn}{1745-2481}, \urlprefix\url{https://doi.org/10.1038/nphys4302}.

\bibitem[{\citenamefont{Tokura et~al.}(2017)\citenamefont{Tokura, Kawasaki, and Nagaosa}}]{Tokura2017}
\bibinfo{author}{\bibfnamefont{Y.}~\bibnamefont{Tokura}}, \bibinfo{author}{\bibfnamefont{M.}~\bibnamefont{Kawasaki}}, \bibnamefont{and} \bibinfo{author}{\bibfnamefont{N.}~\bibnamefont{Nagaosa}}, \bibinfo{journal}{Nature Physics} \textbf{\bibinfo{volume}{13}}, \bibinfo{pages}{1056} (\bibinfo{year}{2017}), ISSN \bibinfo{issn}{1745-2481}, \urlprefix\url{https://doi.org/10.1038/nphys4274}.

\bibitem[{\citenamefont{Basov et~al.}(2017)\citenamefont{Basov, Averitt, and Hsieh}}]{Basov2017}
\bibinfo{author}{\bibfnamefont{D.~N.} \bibnamefont{Basov}}, \bibinfo{author}{\bibfnamefont{R.~D.} \bibnamefont{Averitt}}, \bibnamefont{and} \bibinfo{author}{\bibfnamefont{D.}~\bibnamefont{Hsieh}}, \bibinfo{journal}{Nature Materials} \textbf{\bibinfo{volume}{16}}, \bibinfo{pages}{1077} (\bibinfo{year}{2017}), ISSN \bibinfo{issn}{1476-4660}, \urlprefix\url{https://doi.org/10.1038/nmat5017}.

\bibitem[{\citenamefont{Nicoletti and Cavalleri}(2016)}]{Nicoletti2016}
\bibinfo{author}{\bibfnamefont{D.}~\bibnamefont{Nicoletti}} \bibnamefont{and} \bibinfo{author}{\bibfnamefont{A.}~\bibnamefont{Cavalleri}}, \bibinfo{journal}{Adv. Opt. Photon.} \textbf{\bibinfo{volume}{8}}, \bibinfo{pages}{401} (\bibinfo{year}{2016}), \urlprefix\url{https://opg.optica.org/aop/abstract.cfm?URI=aop-8-3-401}.

\bibitem[{\citenamefont{Yang et~al.}(2023)\citenamefont{Yang, Li, Fiebig, and Pal}}]{Yang2023}
\bibinfo{author}{\bibfnamefont{C.-J.} \bibnamefont{Yang}}, \bibinfo{author}{\bibfnamefont{J.}~\bibnamefont{Li}}, \bibinfo{author}{\bibfnamefont{M.}~\bibnamefont{Fiebig}}, \bibnamefont{and} \bibinfo{author}{\bibfnamefont{S.}~\bibnamefont{Pal}}, \bibinfo{journal}{Nature Reviews Materials} \textbf{\bibinfo{volume}{8}}, \bibinfo{pages}{518} (\bibinfo{year}{2023}), ISSN \bibinfo{issn}{2058-8437}, \urlprefix\url{https://doi.org/10.1038/s41578-023-00566-w}.

\bibitem[{\citenamefont{Davies and Linfield}(2004)}]{Davies2004}
\bibinfo{author}{\bibfnamefont{G.}~\bibnamefont{Davies}} \bibnamefont{and} \bibinfo{author}{\bibfnamefont{E.}~\bibnamefont{Linfield}}, \bibinfo{journal}{Physics World} \textbf{\bibinfo{volume}{17}}, \bibinfo{pages}{37} (\bibinfo{year}{2004}), \urlprefix\url{https://dx.doi.org/10.1088/2058-7058/17/4/34}.

\bibitem[{\citenamefont{Friebolin}(1991)}]{Friebolin1991}
\bibinfo{author}{\bibfnamefont{H.}~\bibnamefont{Friebolin}}, \emph{\bibinfo{title}{Basic One- and Two-dimensional NMR Spectroscopy}} (\bibinfo{publisher}{VCH}, \bibinfo{year}{1991}), ISBN \bibinfo{isbn}{9783527281084}, \urlprefix\url{https://books.google.ch/books?id=cOtqAAAAMAAJ}.

\bibitem[{\citenamefont{Keeler}(2010)}]{Keeler2010}
\bibinfo{author}{\bibfnamefont{J.}~\bibnamefont{Keeler}}, \emph{\bibinfo{title}{Understanding NMR Spectroscopy}} (\bibinfo{publisher}{Wiley}, \bibinfo{year}{2010}).

\bibitem[{\citenamefont{Hamm and Zanni}(2012)}]{HammZanni2012}
\bibinfo{author}{\bibfnamefont{P.}~\bibnamefont{Hamm}} \bibnamefont{and} \bibinfo{author}{\bibfnamefont{M.}~\bibnamefont{Zanni}}, \emph{\bibinfo{title}{Concepts and Methods of 2D Infrared Spectroscopy}} (\bibinfo{publisher}{Cambridge University Press}, \bibinfo{address}{Cambridge}, \bibinfo{year}{2012}).

\bibitem[{\citenamefont{Li et~al.}(2023)\citenamefont{Li, Lomsadze, Moody, Smallwood, and Cundiff}}]{MDCS_Book}
\bibinfo{author}{\bibfnamefont{H.}~\bibnamefont{Li}}, \bibinfo{author}{\bibfnamefont{B.}~\bibnamefont{Lomsadze}}, \bibinfo{author}{\bibfnamefont{G.}~\bibnamefont{Moody}}, \bibinfo{author}{\bibfnamefont{C.}~\bibnamefont{Smallwood}}, \bibnamefont{and} \bibinfo{author}{\bibfnamefont{S.}~\bibnamefont{Cundiff}}, \emph{\bibinfo{title}{Optical Multidimensional Coherent Spectroscopy}} (\bibinfo{publisher}{Oxford University Press}, \bibinfo{address}{Oxford}, \bibinfo{year}{2023}).

\bibitem[{\citenamefont{Cundiff and Mukamel}(2013)}]{Cundiff2013}
\bibinfo{author}{\bibfnamefont{S.~T.} \bibnamefont{Cundiff}} \bibnamefont{and} \bibinfo{author}{\bibfnamefont{S.}~\bibnamefont{Mukamel}}, \bibinfo{journal}{Physics Today} \textbf{\bibinfo{volume}{66}}, \bibinfo{pages}{44} (\bibinfo{year}{2013}), ISSN \bibinfo{issn}{0031-9228}, \eprint{https://pubs.aip.org/physicstoday/article-pdf/66/7/44/10097716/44\_1\_online.pdf}, \urlprefix\url{https://doi.org/10.1063/PT.3.2047}.

\bibitem[{\citenamefont{Fuller and Ogilvie}(2015)}]{Fuller2015}
\bibinfo{author}{\bibfnamefont{F.~D.} \bibnamefont{Fuller}} \bibnamefont{and} \bibinfo{author}{\bibfnamefont{J.~P.} \bibnamefont{Ogilvie}}, \bibinfo{journal}{Annual Review of Physical Chemistry} \textbf{\bibinfo{volume}{66}}, \bibinfo{pages}{667} (\bibinfo{year}{2015}), ISSN \bibinfo{issn}{1545-1593}, \urlprefix\url{https://www.annualreviews.org/content/journals/10.1146/annurev-physchem-040513-103623}.

\bibitem[{\citenamefont{Liu et~al.}(2022)\citenamefont{Liu, Almeida, Padilha, and Cundiff}}]{Liu_2022}
\bibinfo{author}{\bibfnamefont{A.}~\bibnamefont{Liu}}, \bibinfo{author}{\bibfnamefont{D.~B.} \bibnamefont{Almeida}}, \bibinfo{author}{\bibfnamefont{L.~A.} \bibnamefont{Padilha}}, \bibnamefont{and} \bibinfo{author}{\bibfnamefont{S.~T.} \bibnamefont{Cundiff}}, \bibinfo{journal}{Journal of Physics: Materials} \textbf{\bibinfo{volume}{5}}, \bibinfo{pages}{021002} (\bibinfo{year}{2022}), \urlprefix\url{https://dx.doi.org/10.1088/2515-7639/ac4fa5}.

\bibitem[{\citenamefont{Lu et~al.}(2019)\citenamefont{Lu, Li, Zhang, Hwang, Ofori-Okai, and Nelson}}]{Lu2019}
\bibinfo{author}{\bibfnamefont{J.}~\bibnamefont{Lu}}, \bibinfo{author}{\bibfnamefont{X.}~\bibnamefont{Li}}, \bibinfo{author}{\bibfnamefont{Y.}~\bibnamefont{Zhang}}, \bibinfo{author}{\bibfnamefont{H.~Y.} \bibnamefont{Hwang}}, \bibinfo{author}{\bibfnamefont{B.~K.} \bibnamefont{Ofori-Okai}}, \bibnamefont{and} \bibinfo{author}{\bibfnamefont{K.~A.} \bibnamefont{Nelson}}, \emph{\bibinfo{title}{Two-Dimensional Spectroscopy at Terahertz Frequencies}} (\bibinfo{publisher}{Springer International Publishing}, \bibinfo{address}{Cham}, \bibinfo{year}{2019}), pp. \bibinfo{pages}{275--320}, ISBN \bibinfo{isbn}{978-3-030-02478-9}, \urlprefix\url{https://doi.org/10.1007/978-3-030-02478-9_7}.

\bibitem[{\citenamefont{Kuehn et~al.}(2009)\citenamefont{Kuehn, Reimann, Woerner, and Elsaesser}}]{Kuehn2009}
\bibinfo{author}{\bibfnamefont{W.}~\bibnamefont{Kuehn}}, \bibinfo{author}{\bibfnamefont{K.}~\bibnamefont{Reimann}}, \bibinfo{author}{\bibfnamefont{M.}~\bibnamefont{Woerner}}, \bibnamefont{and} \bibinfo{author}{\bibfnamefont{T.}~\bibnamefont{Elsaesser}}, \bibinfo{journal}{The Journal of Chemical Physics} \textbf{\bibinfo{volume}{130}}, \bibinfo{pages}{164503} (\bibinfo{year}{2009}), ISSN \bibinfo{issn}{0021-9606}, \eprint{https://pubs.aip.org/aip/jcp/article-pdf/doi/10.1063/1.3120766/15649633/164503\_1\_online.pdf}, \urlprefix\url{https://doi.org/10.1063/1.3120766}.

\bibitem[{\citenamefont{Lu et~al.}(2016)\citenamefont{Lu, Zhang, Hwang, Ofori-Okai, Fleischer, and Nelson}}]{Lu2016}
\bibinfo{author}{\bibfnamefont{J.}~\bibnamefont{Lu}}, \bibinfo{author}{\bibfnamefont{Y.}~\bibnamefont{Zhang}}, \bibinfo{author}{\bibfnamefont{H.~Y.} \bibnamefont{Hwang}}, \bibinfo{author}{\bibfnamefont{B.~K.} \bibnamefont{Ofori-Okai}}, \bibinfo{author}{\bibfnamefont{S.}~\bibnamefont{Fleischer}}, \bibnamefont{and} \bibinfo{author}{\bibfnamefont{K.~A.} \bibnamefont{Nelson}}, \bibinfo{journal}{Proceedings of the National Academy of Sciences} \textbf{\bibinfo{volume}{113}}, \bibinfo{pages}{11800} (\bibinfo{year}{2016}), \urlprefix\url{https://doi.org/10.1073/pnas.1609558113}.

\bibitem[{\citenamefont{Maag et~al.}(2016)\citenamefont{Maag, Bayer, Baierl, Hohenleutner, Korn, Sch{\"u}ller, Schuh, Bougeard, Lange, Huber et~al.}}]{Maag2016}
\bibinfo{author}{\bibfnamefont{T.}~\bibnamefont{Maag}}, \bibinfo{author}{\bibfnamefont{A.}~\bibnamefont{Bayer}}, \bibinfo{author}{\bibfnamefont{S.}~\bibnamefont{Baierl}}, \bibinfo{author}{\bibfnamefont{M.}~\bibnamefont{Hohenleutner}}, \bibinfo{author}{\bibfnamefont{T.}~\bibnamefont{Korn}}, \bibinfo{author}{\bibfnamefont{C.}~\bibnamefont{Sch{\"u}ller}}, \bibinfo{author}{\bibfnamefont{D.}~\bibnamefont{Schuh}}, \bibinfo{author}{\bibfnamefont{D.}~\bibnamefont{Bougeard}}, \bibinfo{author}{\bibfnamefont{C.}~\bibnamefont{Lange}}, \bibinfo{author}{\bibfnamefont{R.}~\bibnamefont{Huber}}, \bibnamefont{et~al.}, \bibinfo{journal}{Nature Physics} \textbf{\bibinfo{volume}{12}}, \bibinfo{pages}{119} (\bibinfo{year}{2016}), ISSN \bibinfo{issn}{1745-2481}, \urlprefix\url{https://doi.org/10.1038/nphys3559}.

\bibitem[{\citenamefont{Houver et~al.}(2019)\citenamefont{Houver, Huber, Savoini, Abreu, and Johnson}}]{Houver2019}
\bibinfo{author}{\bibfnamefont{S.}~\bibnamefont{Houver}}, \bibinfo{author}{\bibfnamefont{L.}~\bibnamefont{Huber}}, \bibinfo{author}{\bibfnamefont{M.}~\bibnamefont{Savoini}}, \bibinfo{author}{\bibfnamefont{E.}~\bibnamefont{Abreu}}, \bibnamefont{and} \bibinfo{author}{\bibfnamefont{S.~L.} \bibnamefont{Johnson}}, \bibinfo{journal}{Optics Express} \textbf{\bibinfo{volume}{27}}, \bibinfo{pages}{10854} (\bibinfo{year}{2019}), \urlprefix\url{https://doi.org/10.1364/OE.27.010854}.

\bibitem[{\citenamefont{Mahmood et~al.}(2021)\citenamefont{Mahmood, Chaudhuri, Gopalakrishnan, Nandkishore, and Armitage}}]{Mahmood2021}
\bibinfo{author}{\bibfnamefont{F.}~\bibnamefont{Mahmood}}, \bibinfo{author}{\bibfnamefont{D.}~\bibnamefont{Chaudhuri}}, \bibinfo{author}{\bibfnamefont{S.}~\bibnamefont{Gopalakrishnan}}, \bibinfo{author}{\bibfnamefont{R.}~\bibnamefont{Nandkishore}}, \bibnamefont{and} \bibinfo{author}{\bibfnamefont{N.~P.} \bibnamefont{Armitage}}, \bibinfo{journal}{Nature Physics} \textbf{\bibinfo{volume}{17}}, \bibinfo{pages}{627} (\bibinfo{year}{2021}), ISSN \bibinfo{issn}{1745-2481}, \urlprefix\url{https://doi.org/10.1038/s41567-020-01149-0}.

\bibitem[{\citenamefont{Pal et~al.}(2021)\citenamefont{Pal, Strkalj, Yang, Weber, Trassin, Woerner, and Fiebig}}]{Pal2021}
\bibinfo{author}{\bibfnamefont{S.}~\bibnamefont{Pal}}, \bibinfo{author}{\bibfnamefont{N.}~\bibnamefont{Strkalj}}, \bibinfo{author}{\bibfnamefont{C.-J.} \bibnamefont{Yang}}, \bibinfo{author}{\bibfnamefont{M.~C.} \bibnamefont{Weber}}, \bibinfo{author}{\bibfnamefont{M.}~\bibnamefont{Trassin}}, \bibinfo{author}{\bibfnamefont{M.}~\bibnamefont{Woerner}}, \bibnamefont{and} \bibinfo{author}{\bibfnamefont{M.}~\bibnamefont{Fiebig}}, \bibinfo{journal}{Phys. Rev. X} \textbf{\bibinfo{volume}{11}}, \bibinfo{pages}{021023} (\bibinfo{year}{2021}), \urlprefix\url{https://link.aps.org/doi/10.1103/PhysRevX.11.021023}.

\bibitem[{\citenamefont{Lin et~al.}(2022)\citenamefont{Lin, Mead, and Blake}}]{Lin2022}
\bibinfo{author}{\bibfnamefont{H.-W.} \bibnamefont{Lin}}, \bibinfo{author}{\bibfnamefont{G.}~\bibnamefont{Mead}}, \bibnamefont{and} \bibinfo{author}{\bibfnamefont{G.~A.} \bibnamefont{Blake}}, \bibinfo{journal}{Phys. Rev. Lett.} \textbf{\bibinfo{volume}{129}}, \bibinfo{pages}{207401} (\bibinfo{year}{2022}), \urlprefix\url{https://link.aps.org/doi/10.1103/PhysRevLett.129.207401}.

\bibitem[{\citenamefont{Luo et~al.}(2023)\citenamefont{Luo, Mootz, Kang, Huang, Eom, Lee, Vaswani, Collantes, Hellstrom, Perakis et~al.}}]{Luo2023}
\bibinfo{author}{\bibfnamefont{L.}~\bibnamefont{Luo}}, \bibinfo{author}{\bibfnamefont{M.}~\bibnamefont{Mootz}}, \bibinfo{author}{\bibfnamefont{J.~H.} \bibnamefont{Kang}}, \bibinfo{author}{\bibfnamefont{C.}~\bibnamefont{Huang}}, \bibinfo{author}{\bibfnamefont{K.}~\bibnamefont{Eom}}, \bibinfo{author}{\bibfnamefont{J.~W.} \bibnamefont{Lee}}, \bibinfo{author}{\bibfnamefont{C.}~\bibnamefont{Vaswani}}, \bibinfo{author}{\bibfnamefont{Y.~G.} \bibnamefont{Collantes}}, \bibinfo{author}{\bibfnamefont{E.~E.} \bibnamefont{Hellstrom}}, \bibinfo{author}{\bibfnamefont{I.~E.} \bibnamefont{Perakis}}, \bibnamefont{et~al.}, \bibinfo{journal}{Nature Physics} \textbf{\bibinfo{volume}{19}}, \bibinfo{pages}{201} (\bibinfo{year}{2023}), ISSN \bibinfo{issn}{1745-2481}, \urlprefix\url{https://doi.org/10.1038/s41567-022-01827-1}.

\bibitem[{\citenamefont{Blank et~al.}(2023)\citenamefont{Blank, Grishunin, Zvezdin, Hai, Wu, Su, Huang, Zvezdin, and Kimel}}]{Blank2023}
\bibinfo{author}{\bibfnamefont{T.~G.~H.} \bibnamefont{Blank}}, \bibinfo{author}{\bibfnamefont{K.~A.} \bibnamefont{Grishunin}}, \bibinfo{author}{\bibfnamefont{K.~A.} \bibnamefont{Zvezdin}}, \bibinfo{author}{\bibfnamefont{N.~T.} \bibnamefont{Hai}}, \bibinfo{author}{\bibfnamefont{J.~C.} \bibnamefont{Wu}}, \bibinfo{author}{\bibfnamefont{S.-H.} \bibnamefont{Su}}, \bibinfo{author}{\bibfnamefont{J.-C.~A.} \bibnamefont{Huang}}, \bibinfo{author}{\bibfnamefont{A.~K.} \bibnamefont{Zvezdin}}, \bibnamefont{and} \bibinfo{author}{\bibfnamefont{A.~V.} \bibnamefont{Kimel}}, \bibinfo{journal}{Phys. Rev. Lett.} \textbf{\bibinfo{volume}{131}}, \bibinfo{pages}{026902} (\bibinfo{year}{2023}), \urlprefix\url{https://link.aps.org/doi/10.1103/PhysRevLett.131.026902}.

\bibitem[{\citenamefont{Liu et~al.}(2023)\citenamefont{Liu, Pavicevic, Michael, Salvador, Dolgirev, Fechner, Disa, Lozano, Li, Gu et~al.}}]{Liu_2023_echo}
\bibinfo{author}{\bibfnamefont{A.}~\bibnamefont{Liu}}, \bibinfo{author}{\bibfnamefont{D.}~\bibnamefont{Pavicevic}}, \bibinfo{author}{\bibfnamefont{M.~H.} \bibnamefont{Michael}}, \bibinfo{author}{\bibfnamefont{A.~G.} \bibnamefont{Salvador}}, \bibinfo{author}{\bibfnamefont{P.~E.} \bibnamefont{Dolgirev}}, \bibinfo{author}{\bibfnamefont{M.}~\bibnamefont{Fechner}}, \bibinfo{author}{\bibfnamefont{A.~S.} \bibnamefont{Disa}}, \bibinfo{author}{\bibfnamefont{P.~M.} \bibnamefont{Lozano}}, \bibinfo{author}{\bibfnamefont{Q.}~\bibnamefont{Li}}, \bibinfo{author}{\bibfnamefont{G.~D.} \bibnamefont{Gu}}, \bibnamefont{et~al.}, \emph{\bibinfo{title}{Probing inhomogeneous cuprate superconductivity by terahertz josephson echo spectroscopy}} (\bibinfo{year}{2023}), \eprint{2308.14849}.

\bibitem[{\citenamefont{Zhang et~al.}(2024{\natexlab{a}})\citenamefont{Zhang, Gao, Chien, Liu, Curtis, Sung, Ma, Ren, Cao, Narang et~al.}}]{Zhang2024}
\bibinfo{author}{\bibfnamefont{Z.}~\bibnamefont{Zhang}}, \bibinfo{author}{\bibfnamefont{F.~Y.} \bibnamefont{Gao}}, \bibinfo{author}{\bibfnamefont{Y.-C.} \bibnamefont{Chien}}, \bibinfo{author}{\bibfnamefont{Z.-J.} \bibnamefont{Liu}}, \bibinfo{author}{\bibfnamefont{J.~B.} \bibnamefont{Curtis}}, \bibinfo{author}{\bibfnamefont{E.~R.} \bibnamefont{Sung}}, \bibinfo{author}{\bibfnamefont{X.}~\bibnamefont{Ma}}, \bibinfo{author}{\bibfnamefont{W.}~\bibnamefont{Ren}}, \bibinfo{author}{\bibfnamefont{S.}~\bibnamefont{Cao}}, \bibinfo{author}{\bibfnamefont{P.}~\bibnamefont{Narang}}, \bibnamefont{et~al.}, \bibinfo{journal}{Nature Physics}  (\bibinfo{year}{2024}{\natexlab{a}}), ISSN \bibinfo{issn}{1745-2481}, \urlprefix\url{https://doi.org/10.1038/s41567-023-02350-7}.

\bibitem[{\citenamefont{Katsumi et~al.}(2024)\citenamefont{Katsumi, Fiore, Udina, Romero, Barbalas, Jesudasan, Raychaudhuri, Seibold, Benfatto, and Armitage}}]{Katsumi2024}
\bibinfo{author}{\bibfnamefont{K.}~\bibnamefont{Katsumi}}, \bibinfo{author}{\bibfnamefont{J.}~\bibnamefont{Fiore}}, \bibinfo{author}{\bibfnamefont{M.}~\bibnamefont{Udina}}, \bibinfo{author}{\bibfnamefont{R.}~\bibnamefont{Romero}}, \bibinfo{author}{\bibfnamefont{D.}~\bibnamefont{Barbalas}}, \bibinfo{author}{\bibfnamefont{J.}~\bibnamefont{Jesudasan}}, \bibinfo{author}{\bibfnamefont{P.}~\bibnamefont{Raychaudhuri}}, \bibinfo{author}{\bibfnamefont{G.}~\bibnamefont{Seibold}}, \bibinfo{author}{\bibfnamefont{L.}~\bibnamefont{Benfatto}}, \bibnamefont{and} \bibinfo{author}{\bibfnamefont{N.~P.} \bibnamefont{Armitage}}, \bibinfo{journal}{Phys. Rev. Lett.} \textbf{\bibinfo{volume}{132}}, \bibinfo{pages}{256903} (\bibinfo{year}{2024}), \urlprefix\url{https://link.aps.org/doi/10.1103/PhysRevLett.132.256903}.

\bibitem[{\citenamefont{Bao et~al.}(2022)\citenamefont{Bao, Tang, Sun, and Zhou}}]{Bao2022}
\bibinfo{author}{\bibfnamefont{C.}~\bibnamefont{Bao}}, \bibinfo{author}{\bibfnamefont{P.}~\bibnamefont{Tang}}, \bibinfo{author}{\bibfnamefont{D.}~\bibnamefont{Sun}}, \bibnamefont{and} \bibinfo{author}{\bibfnamefont{S.}~\bibnamefont{Zhou}}, \bibinfo{journal}{Nature Reviews Physics} \textbf{\bibinfo{volume}{4}}, \bibinfo{pages}{33} (\bibinfo{year}{2022}), ISSN \bibinfo{issn}{2522-5820}, \urlprefix\url{https://doi.org/10.1038/s42254-021-00388-1}.

\bibitem[{\citenamefont{Gabriele et~al.}(2022)\citenamefont{Gabriele, Castellani, and Benfatto}}]{Gabriele2022}
\bibinfo{author}{\bibfnamefont{F.}~\bibnamefont{Gabriele}}, \bibinfo{author}{\bibfnamefont{C.}~\bibnamefont{Castellani}}, \bibnamefont{and} \bibinfo{author}{\bibfnamefont{L.}~\bibnamefont{Benfatto}}, \bibinfo{journal}{Phys. Rev. Res.} \textbf{\bibinfo{volume}{4}}, \bibinfo{pages}{023112} (\bibinfo{year}{2022}), \urlprefix\url{https://link.aps.org/doi/10.1103/PhysRevResearch.4.023112}.

\bibitem[{\citenamefont{Fiore et~al.}(2024)\citenamefont{Fiore, Sellati, Gabriele, Castellani, Seibold, Udina, and Benfatto}}]{Fiore2023}
\bibinfo{author}{\bibfnamefont{J.}~\bibnamefont{Fiore}}, \bibinfo{author}{\bibfnamefont{N.}~\bibnamefont{Sellati}}, \bibinfo{author}{\bibfnamefont{F.}~\bibnamefont{Gabriele}}, \bibinfo{author}{\bibfnamefont{C.}~\bibnamefont{Castellani}}, \bibinfo{author}{\bibfnamefont{G.}~\bibnamefont{Seibold}}, \bibinfo{author}{\bibfnamefont{M.}~\bibnamefont{Udina}}, \bibnamefont{and} \bibinfo{author}{\bibfnamefont{L.}~\bibnamefont{Benfatto}}, \bibinfo{journal}{Phys. Rev. B} \textbf{\bibinfo{volume}{110}}, \bibinfo{pages}{L060504} (\bibinfo{year}{2024}), \urlprefix\url{https://link.aps.org/doi/10.1103/PhysRevB.110.L060504}.

\bibitem[{\citenamefont{Salvador et~al.}(2024)\citenamefont{Salvador, Dolgirev, Michael, Liu, Pavicevic, Fechner, Cavalleri, and Demler}}]{Salvador2024}
\bibinfo{author}{\bibfnamefont{A.~G.} \bibnamefont{Salvador}}, \bibinfo{author}{\bibfnamefont{P.~E.} \bibnamefont{Dolgirev}}, \bibinfo{author}{\bibfnamefont{M.~H.} \bibnamefont{Michael}}, \bibinfo{author}{\bibfnamefont{A.}~\bibnamefont{Liu}}, \bibinfo{author}{\bibfnamefont{D.}~\bibnamefont{Pavicevic}}, \bibinfo{author}{\bibfnamefont{M.}~\bibnamefont{Fechner}}, \bibinfo{author}{\bibfnamefont{A.}~\bibnamefont{Cavalleri}}, \bibnamefont{and} \bibinfo{author}{\bibfnamefont{E.}~\bibnamefont{Demler}}, \emph{\bibinfo{title}{Principles of 2d terahertz spectroscopy of collective excitations: the case of josephson plasmons in layered superconductors}} (\bibinfo{year}{2024}), \eprint{2401.05503}.

\bibitem[{\citenamefont{Strogatz}(2024)}]{Strogatz2024_Book}
\bibinfo{author}{\bibfnamefont{S.}~\bibnamefont{Strogatz}}, \emph{\bibinfo{title}{Nonlinear Dynamics and Chaos}} (\bibinfo{publisher}{CRC Press}, \bibinfo{address}{Boca Raton}, \bibinfo{year}{2024}).

\bibitem[{\citenamefont{Liu and Disa}(2024)}]{Liu2024_OptExpress}
\bibinfo{author}{\bibfnamefont{A.}~\bibnamefont{Liu}} \bibnamefont{and} \bibinfo{author}{\bibfnamefont{A.}~\bibnamefont{Disa}}, \bibinfo{journal}{Opt. Express} \textbf{\bibinfo{volume}{32}}, \bibinfo{pages}{28160} (\bibinfo{year}{2024}), \urlprefix\url{https://opg.optica.org/oe/abstract.cfm?URI=oe-32-16-28160}.

\bibitem[{\citenamefont{Rajasekaran et~al.}(2016)\citenamefont{Rajasekaran, Casandruc, Laplace, Nicoletti, Gu, Clark, Jaksch, and Cavalleri}}]{Rajasekaran2016}
\bibinfo{author}{\bibfnamefont{S.}~\bibnamefont{Rajasekaran}}, \bibinfo{author}{\bibfnamefont{E.}~\bibnamefont{Casandruc}}, \bibinfo{author}{\bibfnamefont{Y.}~\bibnamefont{Laplace}}, \bibinfo{author}{\bibfnamefont{D.}~\bibnamefont{Nicoletti}}, \bibinfo{author}{\bibfnamefont{G.~D.} \bibnamefont{Gu}}, \bibinfo{author}{\bibfnamefont{S.~R.} \bibnamefont{Clark}}, \bibinfo{author}{\bibfnamefont{D.}~\bibnamefont{Jaksch}}, \bibnamefont{and} \bibinfo{author}{\bibfnamefont{A.}~\bibnamefont{Cavalleri}}, \bibinfo{journal}{Nature Physics} \textbf{\bibinfo{volume}{12}}, \bibinfo{pages}{1012} (\bibinfo{year}{2016}), ISSN \bibinfo{issn}{1745-2481}, \urlprefix\url{https://doi.org/10.1038/nphys3819}.

\bibitem[{\citenamefont{Kim et~al.}(2009)\citenamefont{Kim, Mukamel, and Scholes}}]{Kim2009}
\bibinfo{author}{\bibfnamefont{J.}~\bibnamefont{Kim}}, \bibinfo{author}{\bibfnamefont{S.}~\bibnamefont{Mukamel}}, \bibnamefont{and} \bibinfo{author}{\bibfnamefont{G.~D.} \bibnamefont{Scholes}}, \bibinfo{journal}{Accounts of Chemical Research} \textbf{\bibinfo{volume}{42}}, \bibinfo{pages}{1375} (\bibinfo{year}{2009}), ISSN \bibinfo{issn}{0001-4842}, \urlprefix\url{https://doi.org/10.1021/ar9000795}.

\bibitem[{\citenamefont{Lu et~al.}(2017)\citenamefont{Lu, Li, Hwang, Ofori-Okai, Kurihara, Suemoto, and Nelson}}]{Lu2017}
\bibinfo{author}{\bibfnamefont{J.}~\bibnamefont{Lu}}, \bibinfo{author}{\bibfnamefont{X.}~\bibnamefont{Li}}, \bibinfo{author}{\bibfnamefont{H.~Y.} \bibnamefont{Hwang}}, \bibinfo{author}{\bibfnamefont{B.~K.} \bibnamefont{Ofori-Okai}}, \bibinfo{author}{\bibfnamefont{T.}~\bibnamefont{Kurihara}}, \bibinfo{author}{\bibfnamefont{T.}~\bibnamefont{Suemoto}}, \bibnamefont{and} \bibinfo{author}{\bibfnamefont{K.~A.} \bibnamefont{Nelson}}, \bibinfo{journal}{Phys. Rev. Lett.} \textbf{\bibinfo{volume}{118}}, \bibinfo{pages}{207204} (\bibinfo{year}{2017}), \urlprefix\url{https://link.aps.org/doi/10.1103/PhysRevLett.118.207204}.

\bibitem[{\citenamefont{Huang et~al.}(2024)\citenamefont{Huang, Luo, Mootz, Shang, Man, Su, Perakis, Yao, Wu, and Wang}}]{Huang2024}
\bibinfo{author}{\bibfnamefont{C.}~\bibnamefont{Huang}}, \bibinfo{author}{\bibfnamefont{L.}~\bibnamefont{Luo}}, \bibinfo{author}{\bibfnamefont{M.}~\bibnamefont{Mootz}}, \bibinfo{author}{\bibfnamefont{J.}~\bibnamefont{Shang}}, \bibinfo{author}{\bibfnamefont{P.}~\bibnamefont{Man}}, \bibinfo{author}{\bibfnamefont{L.}~\bibnamefont{Su}}, \bibinfo{author}{\bibfnamefont{I.~E.} \bibnamefont{Perakis}}, \bibinfo{author}{\bibfnamefont{Y.~X.} \bibnamefont{Yao}}, \bibinfo{author}{\bibfnamefont{A.}~\bibnamefont{Wu}}, \bibnamefont{and} \bibinfo{author}{\bibfnamefont{J.}~\bibnamefont{Wang}}, \bibinfo{journal}{Nature Communications} \textbf{\bibinfo{volume}{15}}, \bibinfo{pages}{3214} (\bibinfo{year}{2024}), ISSN \bibinfo{issn}{2041-1723}, \urlprefix\url{https://doi.org/10.1038/s41467-024-47471-6}.

\bibitem[{\citenamefont{Fulmer et~al.}(2004)\citenamefont{Fulmer, Mukherjee, Krummel, and Zanni}}]{Fulmer2004}
\bibinfo{author}{\bibfnamefont{E.~C.} \bibnamefont{Fulmer}}, \bibinfo{author}{\bibfnamefont{P.}~\bibnamefont{Mukherjee}}, \bibinfo{author}{\bibfnamefont{A.~T.} \bibnamefont{Krummel}}, \bibnamefont{and} \bibinfo{author}{\bibfnamefont{M.~T.} \bibnamefont{Zanni}}, \bibinfo{journal}{The Journal of Chemical Physics} \textbf{\bibinfo{volume}{120}}, \bibinfo{pages}{8067} (\bibinfo{year}{2004}), ISSN \bibinfo{issn}{0021-9606}, \eprint{https://pubs.aip.org/aip/jcp/article-pdf/120/17/8067/19134111/8067\_1\_online.pdf}, \urlprefix\url{https://doi.org/10.1063/1.1649725}.

\bibitem[{\citenamefont{Wortis}(1963)}]{Wortis1963}
\bibinfo{author}{\bibfnamefont{M.}~\bibnamefont{Wortis}}, \bibinfo{journal}{Phys. Rev.} \textbf{\bibinfo{volume}{132}}, \bibinfo{pages}{85} (\bibinfo{year}{1963}), \urlprefix\url{https://link.aps.org/doi/10.1103/PhysRev.132.85}.

\bibitem[{\citenamefont{Cohen and Ruvalds}(1969)}]{Cohen1969}
\bibinfo{author}{\bibfnamefont{M.~H.} \bibnamefont{Cohen}} \bibnamefont{and} \bibinfo{author}{\bibfnamefont{J.}~\bibnamefont{Ruvalds}}, \bibinfo{journal}{Phys. Rev. Lett.} \textbf{\bibinfo{volume}{23}}, \bibinfo{pages}{1378} (\bibinfo{year}{1969}), \urlprefix\url{https://link.aps.org/doi/10.1103/PhysRevLett.23.1378}.

\bibitem[{\citenamefont{Khalil et~al.}(2003)\citenamefont{Khalil, Demird{\"o}ven, and Tokmakoff}}]{Khalil2003}
\bibinfo{author}{\bibfnamefont{M.}~\bibnamefont{Khalil}}, \bibinfo{author}{\bibfnamefont{N.}~\bibnamefont{Demird{\"o}ven}}, \bibnamefont{and} \bibinfo{author}{\bibfnamefont{A.}~\bibnamefont{Tokmakoff}}, \bibinfo{journal}{The Journal of Physical Chemistry A} \textbf{\bibinfo{volume}{107}}, \bibinfo{pages}{5258} (\bibinfo{year}{2003}), ISSN \bibinfo{issn}{1089-5639}, \urlprefix\url{https://doi.org/10.1021/jp0219247}.

\bibitem[{\citenamefont{Gabriele et~al.}(2021)\citenamefont{Gabriele, Udina, and Benfatto}}]{Gabriele2021}
\bibinfo{author}{\bibfnamefont{F.}~\bibnamefont{Gabriele}}, \bibinfo{author}{\bibfnamefont{M.}~\bibnamefont{Udina}}, \bibnamefont{and} \bibinfo{author}{\bibfnamefont{L.}~\bibnamefont{Benfatto}}, \bibinfo{journal}{Nature Communications} \textbf{\bibinfo{volume}{12}}, \bibinfo{pages}{752} (\bibinfo{year}{2021}), ISSN \bibinfo{issn}{2041-1723}, \urlprefix\url{https://doi.org/10.1038/s41467-021-21041-6}.

\bibitem[{\citenamefont{Klemens}(1966)}]{Klemens1966}
\bibinfo{author}{\bibfnamefont{P.~G.} \bibnamefont{Klemens}}, \bibinfo{journal}{Phys. Rev.} \textbf{\bibinfo{volume}{148}}, \bibinfo{pages}{845} (\bibinfo{year}{1966}), \urlprefix\url{https://link.aps.org/doi/10.1103/PhysRev.148.845}.

\bibitem[{\citenamefont{Pirro et~al.}(2021)\citenamefont{Pirro, Vasyuchka, Serga, and Hillebrands}}]{Pirro2021}
\bibinfo{author}{\bibfnamefont{P.}~\bibnamefont{Pirro}}, \bibinfo{author}{\bibfnamefont{V.~I.} \bibnamefont{Vasyuchka}}, \bibinfo{author}{\bibfnamefont{A.~A.} \bibnamefont{Serga}}, \bibnamefont{and} \bibinfo{author}{\bibfnamefont{B.}~\bibnamefont{Hillebrands}}, \bibinfo{journal}{Nature Reviews Materials} \textbf{\bibinfo{volume}{6}}, \bibinfo{pages}{1114} (\bibinfo{year}{2021}), ISSN \bibinfo{issn}{2058-8437}, \urlprefix\url{https://doi.org/10.1038/s41578-021-00332-w}.

\bibitem[{\citenamefont{Kim et~al.}(2024)\citenamefont{Kim, Kovalev, Udina, Haenel, Kim, Puviani, Cristiani, Ilyakov, de~Oliveira, Ponomaryov et~al.}}]{Kim2024}
\bibinfo{author}{\bibfnamefont{M.-J.} \bibnamefont{Kim}}, \bibinfo{author}{\bibfnamefont{S.}~\bibnamefont{Kovalev}}, \bibinfo{author}{\bibfnamefont{M.}~\bibnamefont{Udina}}, \bibinfo{author}{\bibfnamefont{R.}~\bibnamefont{Haenel}}, \bibinfo{author}{\bibfnamefont{G.}~\bibnamefont{Kim}}, \bibinfo{author}{\bibfnamefont{M.}~\bibnamefont{Puviani}}, \bibinfo{author}{\bibfnamefont{G.}~\bibnamefont{Cristiani}}, \bibinfo{author}{\bibfnamefont{I.}~\bibnamefont{Ilyakov}}, \bibinfo{author}{\bibfnamefont{T.~V. A.~G.} \bibnamefont{de~Oliveira}}, \bibinfo{author}{\bibfnamefont{A.}~\bibnamefont{Ponomaryov}}, \bibnamefont{et~al.}, \bibinfo{journal}{Science Advances} \textbf{\bibinfo{volume}{10}}, \bibinfo{pages}{eadi7598} (\bibinfo{year}{2024}), \urlprefix\url{https://doi.org/10.1126/sciadv.adi7598}.

\bibitem[{\citenamefont{Puviani et~al.}(2023)\citenamefont{Puviani, Haenel, and Manske}}]{Puviani2023}
\bibinfo{author}{\bibfnamefont{M.}~\bibnamefont{Puviani}}, \bibinfo{author}{\bibfnamefont{R.}~\bibnamefont{Haenel}}, \bibnamefont{and} \bibinfo{author}{\bibfnamefont{D.}~\bibnamefont{Manske}}, \bibinfo{journal}{Phys. Rev. B} \textbf{\bibinfo{volume}{107}}, \bibinfo{pages}{094501} (\bibinfo{year}{2023}), \urlprefix\url{https://link.aps.org/doi/10.1103/PhysRevB.107.094501}.

\bibitem[{\citenamefont{Folpini et~al.}(2017)\citenamefont{Folpini, Reimann, Woerner, Elsaesser, Hoja, and Tkatchenko}}]{Folpini2017}
\bibinfo{author}{\bibfnamefont{G.}~\bibnamefont{Folpini}}, \bibinfo{author}{\bibfnamefont{K.}~\bibnamefont{Reimann}}, \bibinfo{author}{\bibfnamefont{M.}~\bibnamefont{Woerner}}, \bibinfo{author}{\bibfnamefont{T.}~\bibnamefont{Elsaesser}}, \bibinfo{author}{\bibfnamefont{J.}~\bibnamefont{Hoja}}, \bibnamefont{and} \bibinfo{author}{\bibfnamefont{A.}~\bibnamefont{Tkatchenko}}, \bibinfo{journal}{Phys. Rev. Lett.} \textbf{\bibinfo{volume}{119}}, \bibinfo{pages}{097404} (\bibinfo{year}{2017}), \urlprefix\url{https://link.aps.org/doi/10.1103/PhysRevLett.119.097404}.

\bibitem[{\citenamefont{Barbalas et~al.}(2023)\citenamefont{Barbalas, au2, Chaudhuri, Mahmood, Nair, Schreiber, Schlom, Shen, and Armitage}}]{Barbalas2023}
\bibinfo{author}{\bibfnamefont{D.}~\bibnamefont{Barbalas}}, \bibinfo{author}{\bibfnamefont{R.~R.~I.} \bibnamefont{au2}}, \bibinfo{author}{\bibfnamefont{D.}~\bibnamefont{Chaudhuri}}, \bibinfo{author}{\bibfnamefont{F.}~\bibnamefont{Mahmood}}, \bibinfo{author}{\bibfnamefont{H.~P.} \bibnamefont{Nair}}, \bibinfo{author}{\bibfnamefont{N.~J.} \bibnamefont{Schreiber}}, \bibinfo{author}{\bibfnamefont{D.~G.} \bibnamefont{Schlom}}, \bibinfo{author}{\bibfnamefont{K.~M.} \bibnamefont{Shen}}, \bibnamefont{and} \bibinfo{author}{\bibfnamefont{N.~P.} \bibnamefont{Armitage}}, \emph{\bibinfo{title}{Energy relaxation and dynamics in the correlated metal sr$_2$ruo$_4$ via thz two-dimensional coherent spectroscopy}} (\bibinfo{year}{2023}), \eprint{2312.13502}, \urlprefix\url{https://arxiv.org/abs/2312.13502}.

\bibitem[{\citenamefont{Abella et~al.}(1966)\citenamefont{Abella, Kurnit, and Hartmann}}]{Abella1966}
\bibinfo{author}{\bibfnamefont{I.~D.} \bibnamefont{Abella}}, \bibinfo{author}{\bibfnamefont{N.~A.} \bibnamefont{Kurnit}}, \bibnamefont{and} \bibinfo{author}{\bibfnamefont{S.~R.} \bibnamefont{Hartmann}}, \bibinfo{journal}{Phys. Rev.} \textbf{\bibinfo{volume}{141}}, \bibinfo{pages}{391} (\bibinfo{year}{1966}), \urlprefix\url{https://link.aps.org/doi/10.1103/PhysRev.141.391}.

\bibitem[{\citenamefont{Hahn}(1950)}]{Hahn1950}
\bibinfo{author}{\bibfnamefont{E.~L.} \bibnamefont{Hahn}}, \bibinfo{journal}{Phys. Rev.} \textbf{\bibinfo{volume}{80}}, \bibinfo{pages}{580} (\bibinfo{year}{1950}), \urlprefix\url{https://link.aps.org/doi/10.1103/PhysRev.80.580}.

\bibitem[{\citenamefont{Siemens et~al.}(2010)\citenamefont{Siemens, Moody, Li, Bristow, and Cundiff}}]{Siemens2010}
\bibinfo{author}{\bibfnamefont{M.~E.} \bibnamefont{Siemens}}, \bibinfo{author}{\bibfnamefont{G.}~\bibnamefont{Moody}}, \bibinfo{author}{\bibfnamefont{H.}~\bibnamefont{Li}}, \bibinfo{author}{\bibfnamefont{A.~D.} \bibnamefont{Bristow}}, \bibnamefont{and} \bibinfo{author}{\bibfnamefont{S.~T.} \bibnamefont{Cundiff}}, \bibinfo{journal}{Opt. Express} \textbf{\bibinfo{volume}{18}}, \bibinfo{pages}{17699} (\bibinfo{year}{2010}), \urlprefix\url{https://opg.optica.org/oe/abstract.cfm?URI=oe-18-17-17699}.

\bibitem[{\citenamefont{Liu et~al.}(2021)\citenamefont{Liu, Cundiff, Almeida, and Ulbricht}}]{Liu2021_MQT}
\bibinfo{author}{\bibfnamefont{A.}~\bibnamefont{Liu}}, \bibinfo{author}{\bibfnamefont{S.~T.} \bibnamefont{Cundiff}}, \bibinfo{author}{\bibfnamefont{D.~B.} \bibnamefont{Almeida}}, \bibnamefont{and} \bibinfo{author}{\bibfnamefont{R.}~\bibnamefont{Ulbricht}}, \bibinfo{journal}{Materials for Quantum Technology} \textbf{\bibinfo{volume}{1}}, \bibinfo{pages}{025002} (\bibinfo{year}{2021}), \urlprefix\url{https://dx.doi.org/10.1088/2633-4356/abf330}.

\bibitem[{\citenamefont{Wan and Armitage}(2019)}]{Wan2019}
\bibinfo{author}{\bibfnamefont{Y.}~\bibnamefont{Wan}} \bibnamefont{and} \bibinfo{author}{\bibfnamefont{N.~P.} \bibnamefont{Armitage}}, \bibinfo{journal}{Phys. Rev. Lett.} \textbf{\bibinfo{volume}{122}}, \bibinfo{pages}{257401} (\bibinfo{year}{2019}), \urlprefix\url{https://link.aps.org/doi/10.1103/PhysRevLett.122.257401}.

\bibitem[{\citenamefont{Li et~al.}(2021)\citenamefont{Li, Oshikawa, and Wan}}]{Li2021}
\bibinfo{author}{\bibfnamefont{Z.-L.} \bibnamefont{Li}}, \bibinfo{author}{\bibfnamefont{M.}~\bibnamefont{Oshikawa}}, \bibnamefont{and} \bibinfo{author}{\bibfnamefont{Y.}~\bibnamefont{Wan}}, \bibinfo{journal}{Phys. Rev. X} \textbf{\bibinfo{volume}{11}}, \bibinfo{pages}{031035} (\bibinfo{year}{2021}), \urlprefix\url{https://link.aps.org/doi/10.1103/PhysRevX.11.031035}.

\bibitem[{\citenamefont{von Hoegen et~al.}(2018)\citenamefont{von Hoegen, Mankowsky, Fechner, F{\"o}rst, and Cavalleri}}]{vonHoegen2018}
\bibinfo{author}{\bibfnamefont{A.}~\bibnamefont{von Hoegen}}, \bibinfo{author}{\bibfnamefont{R.}~\bibnamefont{Mankowsky}}, \bibinfo{author}{\bibfnamefont{M.}~\bibnamefont{Fechner}}, \bibinfo{author}{\bibfnamefont{M.}~\bibnamefont{F{\"o}rst}}, \bibnamefont{and} \bibinfo{author}{\bibfnamefont{A.}~\bibnamefont{Cavalleri}}, \bibinfo{journal}{Nature} \textbf{\bibinfo{volume}{555}}, \bibinfo{pages}{79} (\bibinfo{year}{2018}), ISSN \bibinfo{issn}{1476-4687}, \urlprefix\url{https://doi.org/10.1038/nature25484}.

\bibitem[{\citenamefont{F{\"o}rst et~al.}(2015)\citenamefont{F{\"o}rst, Mankowsky, and Cavalleri}}]{Forst2015}
\bibinfo{author}{\bibfnamefont{M.}~\bibnamefont{F{\"o}rst}}, \bibinfo{author}{\bibfnamefont{R.}~\bibnamefont{Mankowsky}}, \bibnamefont{and} \bibinfo{author}{\bibfnamefont{A.}~\bibnamefont{Cavalleri}}, \bibinfo{journal}{Accounts of Chemical Research} \textbf{\bibinfo{volume}{48}}, \bibinfo{pages}{380} (\bibinfo{year}{2015}), ISSN \bibinfo{issn}{0001-4842}, \urlprefix\url{https://doi.org/10.1021/ar500391x}.

\bibitem[{\citenamefont{F{\"o}rst et~al.}(2011)\citenamefont{F{\"o}rst, Manzoni, Kaiser, Tomioka, Tokura, Merlin, and Cavalleri}}]{Forst2011}
\bibinfo{author}{\bibfnamefont{M.}~\bibnamefont{F{\"o}rst}}, \bibinfo{author}{\bibfnamefont{C.}~\bibnamefont{Manzoni}}, \bibinfo{author}{\bibfnamefont{S.}~\bibnamefont{Kaiser}}, \bibinfo{author}{\bibfnamefont{Y.}~\bibnamefont{Tomioka}}, \bibinfo{author}{\bibfnamefont{Y.}~\bibnamefont{Tokura}}, \bibinfo{author}{\bibfnamefont{R.}~\bibnamefont{Merlin}}, \bibnamefont{and} \bibinfo{author}{\bibfnamefont{A.}~\bibnamefont{Cavalleri}}, \bibinfo{journal}{Nature Physics} \textbf{\bibinfo{volume}{7}}, \bibinfo{pages}{854} (\bibinfo{year}{2011}), ISSN \bibinfo{issn}{1745-2481}, \urlprefix\url{https://doi.org/10.1038/nphys2055}.

\bibitem[{\citenamefont{Juraschek et~al.}(2017)\citenamefont{Juraschek, Fechner, and Spaldin}}]{Juraschek2017}
\bibinfo{author}{\bibfnamefont{D.~M.} \bibnamefont{Juraschek}}, \bibinfo{author}{\bibfnamefont{M.}~\bibnamefont{Fechner}}, \bibnamefont{and} \bibinfo{author}{\bibfnamefont{N.~A.} \bibnamefont{Spaldin}}, \bibinfo{journal}{Phys. Rev. Lett.} \textbf{\bibinfo{volume}{118}}, \bibinfo{pages}{054101} (\bibinfo{year}{2017}), \urlprefix\url{https://link.aps.org/doi/10.1103/PhysRevLett.118.054101}.

\bibitem[{\citenamefont{Zhang et~al.}(2024{\natexlab{b}})\citenamefont{Zhang, Gao, Curtis, Liu, Chien, von Hoegen, Wong, Kurihara, Suemoto, Narang et~al.}}]{Zhang2024_2}
\bibinfo{author}{\bibfnamefont{Z.}~\bibnamefont{Zhang}}, \bibinfo{author}{\bibfnamefont{F.~Y.} \bibnamefont{Gao}}, \bibinfo{author}{\bibfnamefont{J.~B.} \bibnamefont{Curtis}}, \bibinfo{author}{\bibfnamefont{Z.-J.} \bibnamefont{Liu}}, \bibinfo{author}{\bibfnamefont{Y.-C.} \bibnamefont{Chien}}, \bibinfo{author}{\bibfnamefont{A.}~\bibnamefont{von Hoegen}}, \bibinfo{author}{\bibfnamefont{M.~T.} \bibnamefont{Wong}}, \bibinfo{author}{\bibfnamefont{T.}~\bibnamefont{Kurihara}}, \bibinfo{author}{\bibfnamefont{T.}~\bibnamefont{Suemoto}}, \bibinfo{author}{\bibfnamefont{P.}~\bibnamefont{Narang}}, \bibnamefont{et~al.}, \bibinfo{journal}{Nature Physics} \textbf{\bibinfo{volume}{20}}, \bibinfo{pages}{801} (\bibinfo{year}{2024}{\natexlab{b}}), ISSN \bibinfo{issn}{1745-2481}, \urlprefix\url{https://doi.org/10.1038/s41567-024-02386-3}.

\bibitem[{\citenamefont{Mashkovich et~al.}(2021)\citenamefont{Mashkovich, Grishunin, Dubrovin, Zvezdin, Pisarev, and Kimel}}]{Mashkovich2021}
\bibinfo{author}{\bibfnamefont{E.~A.} \bibnamefont{Mashkovich}}, \bibinfo{author}{\bibfnamefont{K.~A.} \bibnamefont{Grishunin}}, \bibinfo{author}{\bibfnamefont{R.~M.} \bibnamefont{Dubrovin}}, \bibinfo{author}{\bibfnamefont{A.~K.} \bibnamefont{Zvezdin}}, \bibinfo{author}{\bibfnamefont{R.~V.} \bibnamefont{Pisarev}}, \bibnamefont{and} \bibinfo{author}{\bibfnamefont{A.~V.} \bibnamefont{Kimel}}, \bibinfo{journal}{Science} \textbf{\bibinfo{volume}{374}}, \bibinfo{pages}{1608} (\bibinfo{year}{2021}), \eprint{https://www.science.org/doi/pdf/10.1126/science.abk1121}, \urlprefix\url{https://www.science.org/doi/abs/10.1126/science.abk1121}.

\bibitem[{\citenamefont{Fechner et~al.}(2024)\citenamefont{Fechner, F{\"o}rst, Orenstein, Krapivin, Disa, Buzzi, von Hoegen, de~la Pena, Nguyen, Mankowsky et~al.}}]{Fechner2024}
\bibinfo{author}{\bibfnamefont{M.}~\bibnamefont{Fechner}}, \bibinfo{author}{\bibfnamefont{M.}~\bibnamefont{F{\"o}rst}}, \bibinfo{author}{\bibfnamefont{G.}~\bibnamefont{Orenstein}}, \bibinfo{author}{\bibfnamefont{V.}~\bibnamefont{Krapivin}}, \bibinfo{author}{\bibfnamefont{A.~S.} \bibnamefont{Disa}}, \bibinfo{author}{\bibfnamefont{M.}~\bibnamefont{Buzzi}}, \bibinfo{author}{\bibfnamefont{A.}~\bibnamefont{von Hoegen}}, \bibinfo{author}{\bibfnamefont{G.}~\bibnamefont{de~la Pena}}, \bibinfo{author}{\bibfnamefont{Q.~L.} \bibnamefont{Nguyen}}, \bibinfo{author}{\bibfnamefont{R.}~\bibnamefont{Mankowsky}}, \bibnamefont{et~al.}, \bibinfo{journal}{Nature Materials} \textbf{\bibinfo{volume}{23}}, \bibinfo{pages}{363} (\bibinfo{year}{2024}), ISSN \bibinfo{issn}{1476-4660}, \urlprefix\url{https://doi.org/10.1038/s41563-023-01791-y}.

\bibitem[{\citenamefont{Khalsa et~al.}(2021)\citenamefont{Khalsa, Benedek, and Moses}}]{Khalsa2021}
\bibinfo{author}{\bibfnamefont{G.}~\bibnamefont{Khalsa}}, \bibinfo{author}{\bibfnamefont{N.~A.} \bibnamefont{Benedek}}, \bibnamefont{and} \bibinfo{author}{\bibfnamefont{J.}~\bibnamefont{Moses}}, \bibinfo{journal}{Phys. Rev. X} \textbf{\bibinfo{volume}{11}}, \bibinfo{pages}{021067} (\bibinfo{year}{2021}), \urlprefix\url{https://link.aps.org/doi/10.1103/PhysRevX.11.021067}.

\bibitem[{\citenamefont{Disa et~al.}(2021)\citenamefont{Disa, Nova, and Cavalleri}}]{Disa2021}
\bibinfo{author}{\bibfnamefont{A.~S.} \bibnamefont{Disa}}, \bibinfo{author}{\bibfnamefont{T.~F.} \bibnamefont{Nova}}, \bibnamefont{and} \bibinfo{author}{\bibfnamefont{A.}~\bibnamefont{Cavalleri}}, \bibinfo{journal}{Nature Physics} \textbf{\bibinfo{volume}{17}}, \bibinfo{pages}{1087} (\bibinfo{year}{2021}), ISSN \bibinfo{issn}{1745-2481}, \urlprefix\url{https://doi.org/10.1038/s41567-021-01366-1}.

\bibitem[{\citenamefont{Nova et~al.}(2019)\citenamefont{Nova, Disa, Fechner, and Cavalleri}}]{Nova2019}
\bibinfo{author}{\bibfnamefont{T.~F.} \bibnamefont{Nova}}, \bibinfo{author}{\bibfnamefont{A.~S.} \bibnamefont{Disa}}, \bibinfo{author}{\bibfnamefont{M.}~\bibnamefont{Fechner}}, \bibnamefont{and} \bibinfo{author}{\bibfnamefont{A.}~\bibnamefont{Cavalleri}}, \bibinfo{journal}{Science} \textbf{\bibinfo{volume}{364}}, \bibinfo{pages}{1075} (\bibinfo{year}{2019}), \eprint{https://www.science.org/doi/pdf/10.1126/science.aaw4911}, \urlprefix\url{https://www.science.org/doi/abs/10.1126/science.aaw4911}.

\bibitem[{\citenamefont{Li et~al.}(2019)\citenamefont{Li, Qiu, Zhang, Baldini, Lu, Rappe, and Nelson}}]{Li2019}
\bibinfo{author}{\bibfnamefont{X.}~\bibnamefont{Li}}, \bibinfo{author}{\bibfnamefont{T.}~\bibnamefont{Qiu}}, \bibinfo{author}{\bibfnamefont{J.}~\bibnamefont{Zhang}}, \bibinfo{author}{\bibfnamefont{E.}~\bibnamefont{Baldini}}, \bibinfo{author}{\bibfnamefont{J.}~\bibnamefont{Lu}}, \bibinfo{author}{\bibfnamefont{A.~M.} \bibnamefont{Rappe}}, \bibnamefont{and} \bibinfo{author}{\bibfnamefont{K.~A.} \bibnamefont{Nelson}}, \bibinfo{journal}{Science} \textbf{\bibinfo{volume}{364}}, \bibinfo{pages}{1079} (\bibinfo{year}{2019}), \eprint{https://www.science.org/doi/pdf/10.1126/science.aaw4913}, \urlprefix\url{https://www.science.org/doi/abs/10.1126/science.aaw4913}.

\bibitem[{\citenamefont{Wang et~al.}(2022)\citenamefont{Wang, Xiao, Park, Zhu, Wang, Taniguchi, Watanabe, Yan, Xiao, Gamelin et~al.}}]{Wang2022}
\bibinfo{author}{\bibfnamefont{X.}~\bibnamefont{Wang}}, \bibinfo{author}{\bibfnamefont{C.}~\bibnamefont{Xiao}}, \bibinfo{author}{\bibfnamefont{H.}~\bibnamefont{Park}}, \bibinfo{author}{\bibfnamefont{J.}~\bibnamefont{Zhu}}, \bibinfo{author}{\bibfnamefont{C.}~\bibnamefont{Wang}}, \bibinfo{author}{\bibfnamefont{T.}~\bibnamefont{Taniguchi}}, \bibinfo{author}{\bibfnamefont{K.}~\bibnamefont{Watanabe}}, \bibinfo{author}{\bibfnamefont{J.}~\bibnamefont{Yan}}, \bibinfo{author}{\bibfnamefont{D.}~\bibnamefont{Xiao}}, \bibinfo{author}{\bibfnamefont{D.~R.} \bibnamefont{Gamelin}}, \bibnamefont{et~al.}, \bibinfo{journal}{Nature} \textbf{\bibinfo{volume}{604}}, \bibinfo{pages}{468} (\bibinfo{year}{2022}), ISSN \bibinfo{issn}{1476-4687}, \urlprefix\url{https://doi.org/10.1038/s41586-022-04472-z}.

\bibitem[{\citenamefont{Disa et~al.}(2023)\citenamefont{Disa, Curtis, Fechner, Liu, von Hoegen, F{\"o}rst, Nova, Narang, Maljuk, Boris et~al.}}]{Disa2023}
\bibinfo{author}{\bibfnamefont{A.~S.} \bibnamefont{Disa}}, \bibinfo{author}{\bibfnamefont{J.}~\bibnamefont{Curtis}}, \bibinfo{author}{\bibfnamefont{M.}~\bibnamefont{Fechner}}, \bibinfo{author}{\bibfnamefont{A.}~\bibnamefont{Liu}}, \bibinfo{author}{\bibfnamefont{A.}~\bibnamefont{von Hoegen}}, \bibinfo{author}{\bibfnamefont{M.}~\bibnamefont{F{\"o}rst}}, \bibinfo{author}{\bibfnamefont{T.~F.} \bibnamefont{Nova}}, \bibinfo{author}{\bibfnamefont{P.}~\bibnamefont{Narang}}, \bibinfo{author}{\bibfnamefont{A.}~\bibnamefont{Maljuk}}, \bibinfo{author}{\bibfnamefont{A.~V.} \bibnamefont{Boris}}, \bibnamefont{et~al.}, \bibinfo{journal}{Nature} \textbf{\bibinfo{volume}{617}}, \bibinfo{pages}{73} (\bibinfo{year}{2023}), ISSN \bibinfo{issn}{1476-4687}, \urlprefix\url{https://doi.org/10.1038/s41586-023-05853-8}.

\bibitem[{\citenamefont{Sie et~al.}(2019)\citenamefont{Sie, Nyby, Pemmaraju, Park, Shen, Yang, Hoffmann, Ofori-Okai, Li, Reid et~al.}}]{Sie2019}
\bibinfo{author}{\bibfnamefont{E.~J.} \bibnamefont{Sie}}, \bibinfo{author}{\bibfnamefont{C.~M.} \bibnamefont{Nyby}}, \bibinfo{author}{\bibfnamefont{C.~D.} \bibnamefont{Pemmaraju}}, \bibinfo{author}{\bibfnamefont{S.~J.} \bibnamefont{Park}}, \bibinfo{author}{\bibfnamefont{X.}~\bibnamefont{Shen}}, \bibinfo{author}{\bibfnamefont{J.}~\bibnamefont{Yang}}, \bibinfo{author}{\bibfnamefont{M.~C.} \bibnamefont{Hoffmann}}, \bibinfo{author}{\bibfnamefont{B.~K.} \bibnamefont{Ofori-Okai}}, \bibinfo{author}{\bibfnamefont{R.}~\bibnamefont{Li}}, \bibinfo{author}{\bibfnamefont{A.~H.} \bibnamefont{Reid}}, \bibnamefont{et~al.}, \bibinfo{journal}{Nature} \textbf{\bibinfo{volume}{565}}, \bibinfo{pages}{61} (\bibinfo{year}{2019}), ISSN \bibinfo{issn}{1476-4687}, \urlprefix\url{https://doi.org/10.1038/s41586-018-0809-4}.

\bibitem[{\citenamefont{Vaswani et~al.}(2020)\citenamefont{Vaswani, Wang, Mudiyanselage, Li, Lozano, Gu, Cheng, Song, Luo, Kim et~al.}}]{Vaswani2020}
\bibinfo{author}{\bibfnamefont{C.}~\bibnamefont{Vaswani}}, \bibinfo{author}{\bibfnamefont{L.-L.} \bibnamefont{Wang}}, \bibinfo{author}{\bibfnamefont{D.~H.} \bibnamefont{Mudiyanselage}}, \bibinfo{author}{\bibfnamefont{Q.}~\bibnamefont{Li}}, \bibinfo{author}{\bibfnamefont{P.~M.} \bibnamefont{Lozano}}, \bibinfo{author}{\bibfnamefont{G.~D.} \bibnamefont{Gu}}, \bibinfo{author}{\bibfnamefont{D.}~\bibnamefont{Cheng}}, \bibinfo{author}{\bibfnamefont{B.}~\bibnamefont{Song}}, \bibinfo{author}{\bibfnamefont{L.}~\bibnamefont{Luo}}, \bibinfo{author}{\bibfnamefont{R.~H.~J.} \bibnamefont{Kim}}, \bibnamefont{et~al.}, \bibinfo{journal}{Phys. Rev. X} \textbf{\bibinfo{volume}{10}}, \bibinfo{pages}{021013} (\bibinfo{year}{2020}), \urlprefix\url{https://link.aps.org/doi/10.1103/PhysRevX.10.021013}.

\bibitem[{\citenamefont{Luo et~al.}(2021)\citenamefont{Luo, Cheng, Song, Wang, Vaswani, Lozano, Gu, Huang, Kim, Liu et~al.}}]{Luo2021}
\bibinfo{author}{\bibfnamefont{L.}~\bibnamefont{Luo}}, \bibinfo{author}{\bibfnamefont{D.}~\bibnamefont{Cheng}}, \bibinfo{author}{\bibfnamefont{B.}~\bibnamefont{Song}}, \bibinfo{author}{\bibfnamefont{L.-L.} \bibnamefont{Wang}}, \bibinfo{author}{\bibfnamefont{C.}~\bibnamefont{Vaswani}}, \bibinfo{author}{\bibfnamefont{P.~M.} \bibnamefont{Lozano}}, \bibinfo{author}{\bibfnamefont{G.}~\bibnamefont{Gu}}, \bibinfo{author}{\bibfnamefont{C.}~\bibnamefont{Huang}}, \bibinfo{author}{\bibfnamefont{R.~H.~J.} \bibnamefont{Kim}}, \bibinfo{author}{\bibfnamefont{Z.}~\bibnamefont{Liu}}, \bibnamefont{et~al.}, \bibinfo{journal}{Nature Materials} \textbf{\bibinfo{volume}{20}}, \bibinfo{pages}{329} (\bibinfo{year}{2021}), ISSN \bibinfo{issn}{1476-4660}, \urlprefix\url{https://doi.org/10.1038/s41563-020-00882-4}.

\bibitem[{\citenamefont{Fausti et~al.}(2011)\citenamefont{Fausti, Tobey, Dean, Kaiser, Dienst, Hoffmann, Pyon, Takayama, Takagi, and Cavalleri}}]{Fausti2011}
\bibinfo{author}{\bibfnamefont{D.}~\bibnamefont{Fausti}}, \bibinfo{author}{\bibfnamefont{R.~I.} \bibnamefont{Tobey}}, \bibinfo{author}{\bibfnamefont{N.}~\bibnamefont{Dean}}, \bibinfo{author}{\bibfnamefont{S.}~\bibnamefont{Kaiser}}, \bibinfo{author}{\bibfnamefont{A.}~\bibnamefont{Dienst}}, \bibinfo{author}{\bibfnamefont{M.~C.} \bibnamefont{Hoffmann}}, \bibinfo{author}{\bibfnamefont{S.}~\bibnamefont{Pyon}}, \bibinfo{author}{\bibfnamefont{T.}~\bibnamefont{Takayama}}, \bibinfo{author}{\bibfnamefont{H.}~\bibnamefont{Takagi}}, \bibnamefont{and} \bibinfo{author}{\bibfnamefont{A.}~\bibnamefont{Cavalleri}}, \bibinfo{journal}{Science} \textbf{\bibinfo{volume}{331}}, \bibinfo{pages}{189} (\bibinfo{year}{2011}), \eprint{https://www.science.org/doi/pdf/10.1126/science.1197294}, \urlprefix\url{https://www.science.org/doi/abs/10.1126/science.1197294}.

\bibitem[{\citenamefont{Mitrano et~al.}(2016)\citenamefont{Mitrano, Cantaluppi, Nicoletti, Kaiser, Perucchi, Lupi, Di~Pietro, Pontiroli, Ricc{\`o}, Clark et~al.}}]{Mitrano2016}
\bibinfo{author}{\bibfnamefont{M.}~\bibnamefont{Mitrano}}, \bibinfo{author}{\bibfnamefont{A.}~\bibnamefont{Cantaluppi}}, \bibinfo{author}{\bibfnamefont{D.}~\bibnamefont{Nicoletti}}, \bibinfo{author}{\bibfnamefont{S.}~\bibnamefont{Kaiser}}, \bibinfo{author}{\bibfnamefont{A.}~\bibnamefont{Perucchi}}, \bibinfo{author}{\bibfnamefont{S.}~\bibnamefont{Lupi}}, \bibinfo{author}{\bibfnamefont{P.}~\bibnamefont{Di~Pietro}}, \bibinfo{author}{\bibfnamefont{D.}~\bibnamefont{Pontiroli}}, \bibinfo{author}{\bibfnamefont{M.}~\bibnamefont{Ricc{\`o}}}, \bibinfo{author}{\bibfnamefont{S.~R.} \bibnamefont{Clark}}, \bibnamefont{et~al.}, \bibinfo{journal}{Nature} \textbf{\bibinfo{volume}{530}}, \bibinfo{pages}{461} (\bibinfo{year}{2016}), ISSN \bibinfo{issn}{1476-4687}, \urlprefix\url{https://doi.org/10.1038/nature16522}.

\bibitem[{\citenamefont{Fava et~al.}(2024)\citenamefont{Fava, De~Vecchi, Jotzu, Buzzi, Gebert, Liu, Keimer, and Cavalleri}}]{Fava2024}
\bibinfo{author}{\bibfnamefont{S.}~\bibnamefont{Fava}}, \bibinfo{author}{\bibfnamefont{G.}~\bibnamefont{De~Vecchi}}, \bibinfo{author}{\bibfnamefont{G.}~\bibnamefont{Jotzu}}, \bibinfo{author}{\bibfnamefont{M.}~\bibnamefont{Buzzi}}, \bibinfo{author}{\bibfnamefont{T.}~\bibnamefont{Gebert}}, \bibinfo{author}{\bibfnamefont{Y.}~\bibnamefont{Liu}}, \bibinfo{author}{\bibfnamefont{B.}~\bibnamefont{Keimer}}, \bibnamefont{and} \bibinfo{author}{\bibfnamefont{A.}~\bibnamefont{Cavalleri}}, \bibinfo{journal}{Nature} \textbf{\bibinfo{volume}{632}}, \bibinfo{pages}{75} (\bibinfo{year}{2024}), ISSN \bibinfo{issn}{1476-4687}, \urlprefix\url{https://doi.org/10.1038/s41586-024-07635-2}.

\bibitem[{\citenamefont{de~la Torre et~al.}(2021)\citenamefont{de~la Torre, Kennes, Claassen, Gerber, McIver, and Sentef}}]{delaTorre2021}
\bibinfo{author}{\bibfnamefont{A.}~\bibnamefont{de~la Torre}}, \bibinfo{author}{\bibfnamefont{D.~M.} \bibnamefont{Kennes}}, \bibinfo{author}{\bibfnamefont{M.}~\bibnamefont{Claassen}}, \bibinfo{author}{\bibfnamefont{S.}~\bibnamefont{Gerber}}, \bibinfo{author}{\bibfnamefont{J.~W.} \bibnamefont{McIver}}, \bibnamefont{and} \bibinfo{author}{\bibfnamefont{M.~A.} \bibnamefont{Sentef}}, \bibinfo{journal}{Rev. Mod. Phys.} \textbf{\bibinfo{volume}{93}}, \bibinfo{pages}{041002} (\bibinfo{year}{2021}), \urlprefix\url{https://link.aps.org/doi/10.1103/RevModPhys.93.041002}.

\bibitem[{\citenamefont{Kogar et~al.}(2020)\citenamefont{Kogar, Zong, Dolgirev, Shen, Straquadine, Bie, Wang, Rohwer, Tung, Yang et~al.}}]{Kogar2020}
\bibinfo{author}{\bibfnamefont{A.}~\bibnamefont{Kogar}}, \bibinfo{author}{\bibfnamefont{A.}~\bibnamefont{Zong}}, \bibinfo{author}{\bibfnamefont{P.~E.} \bibnamefont{Dolgirev}}, \bibinfo{author}{\bibfnamefont{X.}~\bibnamefont{Shen}}, \bibinfo{author}{\bibfnamefont{J.}~\bibnamefont{Straquadine}}, \bibinfo{author}{\bibfnamefont{Y.-Q.} \bibnamefont{Bie}}, \bibinfo{author}{\bibfnamefont{X.}~\bibnamefont{Wang}}, \bibinfo{author}{\bibfnamefont{T.}~\bibnamefont{Rohwer}}, \bibinfo{author}{\bibfnamefont{I.-C.} \bibnamefont{Tung}}, \bibinfo{author}{\bibfnamefont{Y.}~\bibnamefont{Yang}}, \bibnamefont{et~al.}, \bibinfo{journal}{Nature Physics} \textbf{\bibinfo{volume}{16}}, \bibinfo{pages}{159} (\bibinfo{year}{2020}), ISSN \bibinfo{issn}{1745-2481}, \urlprefix\url{https://doi.org/10.1038/s41567-019-0705-3}.

\bibitem[{\citenamefont{Turner and Nelson}(2010)}]{Turner2010}
\bibinfo{author}{\bibfnamefont{D.~B.} \bibnamefont{Turner}} \bibnamefont{and} \bibinfo{author}{\bibfnamefont{K.~A.} \bibnamefont{Nelson}}, \bibinfo{journal}{Nature} \textbf{\bibinfo{volume}{466}}, \bibinfo{pages}{1089} (\bibinfo{year}{2010}), ISSN \bibinfo{issn}{1476-4687}, \urlprefix\url{https://doi.org/10.1038/nature09286}.

\bibitem[{\citenamefont{Halperin}(2019)}]{Halperin2019}
\bibinfo{author}{\bibfnamefont{B.~I.} \bibnamefont{Halperin}}, \bibinfo{journal}{Physics Today} \textbf{\bibinfo{volume}{72}}, \bibinfo{pages}{42} (\bibinfo{year}{2019}), ISSN \bibinfo{issn}{0031-9228}, \eprint{https://pubs.aip.org/physicstoday/article-pdf/72/2/42/10122309/42\_1\_online.pdf}, \urlprefix\url{https://doi.org/10.1063/PT.3.4137}.

\bibitem[{\citenamefont{Taherian et~al.}(2024)\citenamefont{Taherian, Först, Liu, Fechner, Pavicevic, von Hoegen, Rowe, Liu, Nakata, Keimer et~al.}}]{Taherian2024}
\bibinfo{author}{\bibfnamefont{N.}~\bibnamefont{Taherian}}, \bibinfo{author}{\bibfnamefont{M.}~\bibnamefont{Först}}, \bibinfo{author}{\bibfnamefont{A.}~\bibnamefont{Liu}}, \bibinfo{author}{\bibfnamefont{M.}~\bibnamefont{Fechner}}, \bibinfo{author}{\bibfnamefont{D.}~\bibnamefont{Pavicevic}}, \bibinfo{author}{\bibfnamefont{A.}~\bibnamefont{von Hoegen}}, \bibinfo{author}{\bibfnamefont{E.}~\bibnamefont{Rowe}}, \bibinfo{author}{\bibfnamefont{Y.}~\bibnamefont{Liu}}, \bibinfo{author}{\bibfnamefont{S.}~\bibnamefont{Nakata}}, \bibinfo{author}{\bibfnamefont{B.}~\bibnamefont{Keimer}}, \bibnamefont{et~al.}, \emph{\bibinfo{title}{Squeezed josephson plasmons in driven yba$_2$cu$_3$o$_{6+x}$}} (\bibinfo{year}{2024}), \eprint{2401.01115}, \urlprefix\url{https://arxiv.org/abs/2401.01115}.

\bibitem[{\citenamefont{von Hoegen et~al.}(2022)\citenamefont{von Hoegen, Fechner, F\"orst, Taherian, Rowe, Ribak, Porras, Keimer, Michael, Demler et~al.}}]{vonHoegen2022}
\bibinfo{author}{\bibfnamefont{A.}~\bibnamefont{von Hoegen}}, \bibinfo{author}{\bibfnamefont{M.}~\bibnamefont{Fechner}}, \bibinfo{author}{\bibfnamefont{M.}~\bibnamefont{F\"orst}}, \bibinfo{author}{\bibfnamefont{N.}~\bibnamefont{Taherian}}, \bibinfo{author}{\bibfnamefont{E.}~\bibnamefont{Rowe}}, \bibinfo{author}{\bibfnamefont{A.}~\bibnamefont{Ribak}}, \bibinfo{author}{\bibfnamefont{J.}~\bibnamefont{Porras}}, \bibinfo{author}{\bibfnamefont{B.}~\bibnamefont{Keimer}}, \bibinfo{author}{\bibfnamefont{M.}~\bibnamefont{Michael}}, \bibinfo{author}{\bibfnamefont{E.}~\bibnamefont{Demler}}, \bibnamefont{et~al.}, \bibinfo{journal}{Phys. Rev. X} \textbf{\bibinfo{volume}{12}}, \bibinfo{pages}{031008} (\bibinfo{year}{2022}), \urlprefix\url{https://link.aps.org/doi/10.1103/PhysRevX.12.031008}.

\bibitem[{\citenamefont{Buzzi et~al.}(2020)\citenamefont{Buzzi, Nicoletti, Fechner, Tancogne-Dejean, Sentef, Georges, Biesner, Uykur, Dressel, Henderson et~al.}}]{Buzzi2020}
\bibinfo{author}{\bibfnamefont{M.}~\bibnamefont{Buzzi}}, \bibinfo{author}{\bibfnamefont{D.}~\bibnamefont{Nicoletti}}, \bibinfo{author}{\bibfnamefont{M.}~\bibnamefont{Fechner}}, \bibinfo{author}{\bibfnamefont{N.}~\bibnamefont{Tancogne-Dejean}}, \bibinfo{author}{\bibfnamefont{M.~A.} \bibnamefont{Sentef}}, \bibinfo{author}{\bibfnamefont{A.}~\bibnamefont{Georges}}, \bibinfo{author}{\bibfnamefont{T.}~\bibnamefont{Biesner}}, \bibinfo{author}{\bibfnamefont{E.}~\bibnamefont{Uykur}}, \bibinfo{author}{\bibfnamefont{M.}~\bibnamefont{Dressel}}, \bibinfo{author}{\bibfnamefont{A.}~\bibnamefont{Henderson}}, \bibnamefont{et~al.}, \bibinfo{journal}{Phys. Rev. X} \textbf{\bibinfo{volume}{10}}, \bibinfo{pages}{031028} (\bibinfo{year}{2020}), \urlprefix\url{https://link.aps.org/doi/10.1103/PhysRevX.10.031028}.

\bibitem[{\citenamefont{Liu et~al.}(2020)\citenamefont{Liu, F\"orst, Fechner, Nicoletti, Porras, Loew, Keimer, and Cavalleri}}]{Liu2020_PRX}
\bibinfo{author}{\bibfnamefont{B.}~\bibnamefont{Liu}}, \bibinfo{author}{\bibfnamefont{M.}~\bibnamefont{F\"orst}}, \bibinfo{author}{\bibfnamefont{M.}~\bibnamefont{Fechner}}, \bibinfo{author}{\bibfnamefont{D.}~\bibnamefont{Nicoletti}}, \bibinfo{author}{\bibfnamefont{J.}~\bibnamefont{Porras}}, \bibinfo{author}{\bibfnamefont{T.}~\bibnamefont{Loew}}, \bibinfo{author}{\bibfnamefont{B.}~\bibnamefont{Keimer}}, \bibnamefont{and} \bibinfo{author}{\bibfnamefont{A.}~\bibnamefont{Cavalleri}}, \bibinfo{journal}{Phys. Rev. X} \textbf{\bibinfo{volume}{10}}, \bibinfo{pages}{011053} (\bibinfo{year}{2020}), \urlprefix\url{https://link.aps.org/doi/10.1103/PhysRevX.10.011053}.

\bibitem[{\citenamefont{Rowe et~al.}(2023)\citenamefont{Rowe, Yuan, Buzzi, Jotzu, Zhu, Fechner, F{\"o}rst, Liu, Pontiroli, Ricc{\`o} et~al.}}]{Rowe2023}
\bibinfo{author}{\bibfnamefont{E.}~\bibnamefont{Rowe}}, \bibinfo{author}{\bibfnamefont{B.}~\bibnamefont{Yuan}}, \bibinfo{author}{\bibfnamefont{M.}~\bibnamefont{Buzzi}}, \bibinfo{author}{\bibfnamefont{G.}~\bibnamefont{Jotzu}}, \bibinfo{author}{\bibfnamefont{Y.}~\bibnamefont{Zhu}}, \bibinfo{author}{\bibfnamefont{M.}~\bibnamefont{Fechner}}, \bibinfo{author}{\bibfnamefont{M.}~\bibnamefont{F{\"o}rst}}, \bibinfo{author}{\bibfnamefont{B.}~\bibnamefont{Liu}}, \bibinfo{author}{\bibfnamefont{D.}~\bibnamefont{Pontiroli}}, \bibinfo{author}{\bibfnamefont{M.}~\bibnamefont{Ricc{\`o}}}, \bibnamefont{et~al.}, \bibinfo{journal}{Nature Physics} \textbf{\bibinfo{volume}{19}}, \bibinfo{pages}{1821} (\bibinfo{year}{2023}), ISSN \bibinfo{issn}{1745-2481}, \urlprefix\url{https://doi.org/10.1038/s41567-023-02235-9}.

\bibitem[{\citenamefont{Li et~al.}(2006)\citenamefont{Li, Zhang, Borca, and Cundiff}}]{Li2006}
\bibinfo{author}{\bibfnamefont{X.}~\bibnamefont{Li}}, \bibinfo{author}{\bibfnamefont{T.}~\bibnamefont{Zhang}}, \bibinfo{author}{\bibfnamefont{C.~N.} \bibnamefont{Borca}}, \bibnamefont{and} \bibinfo{author}{\bibfnamefont{S.~T.} \bibnamefont{Cundiff}}, \bibinfo{journal}{Phys. Rev. Lett.} \textbf{\bibinfo{volume}{96}}, \bibinfo{pages}{057406} (\bibinfo{year}{2006}), \urlprefix\url{https://link.aps.org/doi/10.1103/PhysRevLett.96.057406}.

\bibitem[{\citenamefont{Ma et~al.}(2021)\citenamefont{Ma, Grushin, and Burch}}]{Ma2021}
\bibinfo{author}{\bibfnamefont{Q.}~\bibnamefont{Ma}}, \bibinfo{author}{\bibfnamefont{A.~G.} \bibnamefont{Grushin}}, \bibnamefont{and} \bibinfo{author}{\bibfnamefont{K.~S.} \bibnamefont{Burch}}, \bibinfo{journal}{Nature Materials} \textbf{\bibinfo{volume}{20}}, \bibinfo{pages}{1601} (\bibinfo{year}{2021}), ISSN \bibinfo{issn}{1476-4660}, \urlprefix\url{https://doi.org/10.1038/s41563-021-00992-7}.

\bibitem[{\citenamefont{Cocker et~al.}(2021)\citenamefont{Cocker, Jelic, Hillenbrand, and Hegmann}}]{Cocker2021}
\bibinfo{author}{\bibfnamefont{T.~L.} \bibnamefont{Cocker}}, \bibinfo{author}{\bibfnamefont{V.}~\bibnamefont{Jelic}}, \bibinfo{author}{\bibfnamefont{R.}~\bibnamefont{Hillenbrand}}, \bibnamefont{and} \bibinfo{author}{\bibfnamefont{F.~A.} \bibnamefont{Hegmann}}, \bibinfo{journal}{Nature Photonics} \textbf{\bibinfo{volume}{15}}, \bibinfo{pages}{558} (\bibinfo{year}{2021}), ISSN \bibinfo{issn}{1749-4893}, \urlprefix\url{https://doi.org/10.1038/s41566-021-00835-6}.

\bibitem[{\citenamefont{Chen et~al.}(2019)\citenamefont{Chen, Hu, Mescall, You, Basov, Dai, and Liu}}]{Chen2019}
\bibinfo{author}{\bibfnamefont{X.}~\bibnamefont{Chen}}, \bibinfo{author}{\bibfnamefont{D.}~\bibnamefont{Hu}}, \bibinfo{author}{\bibfnamefont{R.}~\bibnamefont{Mescall}}, \bibinfo{author}{\bibfnamefont{G.}~\bibnamefont{You}}, \bibinfo{author}{\bibfnamefont{D.~N.} \bibnamefont{Basov}}, \bibinfo{author}{\bibfnamefont{Q.}~\bibnamefont{Dai}}, \bibnamefont{and} \bibinfo{author}{\bibfnamefont{M.}~\bibnamefont{Liu}}, \bibinfo{journal}{Advanced Materials} \textbf{\bibinfo{volume}{31}}, \bibinfo{pages}{1804774} (\bibinfo{year}{2019}), \eprint{https://onlinelibrary.wiley.com/doi/pdf/10.1002/adma.201804774}, \urlprefix\url{https://onlinelibrary.wiley.com/doi/abs/10.1002/adma.201804774}.

\bibitem[{\citenamefont{Vojta}(2019)}]{Vojta2019}
\bibinfo{author}{\bibfnamefont{T.}~\bibnamefont{Vojta}}, \bibinfo{journal}{Annual Review of Condensed Matter Physics} \textbf{\bibinfo{volume}{10}}, \bibinfo{pages}{233} (\bibinfo{year}{2019}), ISSN \bibinfo{issn}{1947-5462}, \urlprefix\url{https://www.annualreviews.org/content/journals/10.1146/annurev-conmatphys-031218-013433}.

\bibitem[{\citenamefont{Gallagher et~al.}(2019)\citenamefont{Gallagher, Yang, Lyu, Tian, Kou, Zhang, Watanabe, Taniguchi, and Wang}}]{Gallagher2019}
\bibinfo{author}{\bibfnamefont{P.}~\bibnamefont{Gallagher}}, \bibinfo{author}{\bibfnamefont{C.-S.} \bibnamefont{Yang}}, \bibinfo{author}{\bibfnamefont{T.}~\bibnamefont{Lyu}}, \bibinfo{author}{\bibfnamefont{F.}~\bibnamefont{Tian}}, \bibinfo{author}{\bibfnamefont{R.}~\bibnamefont{Kou}}, \bibinfo{author}{\bibfnamefont{H.}~\bibnamefont{Zhang}}, \bibinfo{author}{\bibfnamefont{K.}~\bibnamefont{Watanabe}}, \bibinfo{author}{\bibfnamefont{T.}~\bibnamefont{Taniguchi}}, \bibnamefont{and} \bibinfo{author}{\bibfnamefont{F.}~\bibnamefont{Wang}}, \bibinfo{journal}{Science} \textbf{\bibinfo{volume}{364}}, \bibinfo{pages}{158} (\bibinfo{year}{2019}), \eprint{https://www.science.org/doi/pdf/10.1126/science.aat8687}, \urlprefix\url{https://www.science.org/doi/abs/10.1126/science.aat8687}.

\bibitem[{\citenamefont{Wang et~al.}(2023)\citenamefont{Wang, Adelinia, Chavez-Cervantes, Matsuyama, Fechner, Buzzi, Meier, and Cavalleri}}]{Wang2023}
\bibinfo{author}{\bibfnamefont{E.}~\bibnamefont{Wang}}, \bibinfo{author}{\bibfnamefont{J.~D.} \bibnamefont{Adelinia}}, \bibinfo{author}{\bibfnamefont{M.}~\bibnamefont{Chavez-Cervantes}}, \bibinfo{author}{\bibfnamefont{T.}~\bibnamefont{Matsuyama}}, \bibinfo{author}{\bibfnamefont{M.}~\bibnamefont{Fechner}}, \bibinfo{author}{\bibfnamefont{M.}~\bibnamefont{Buzzi}}, \bibinfo{author}{\bibfnamefont{G.}~\bibnamefont{Meier}}, \bibnamefont{and} \bibinfo{author}{\bibfnamefont{A.}~\bibnamefont{Cavalleri}}, \bibinfo{journal}{Nature Communications} \textbf{\bibinfo{volume}{14}}, \bibinfo{pages}{7233} (\bibinfo{year}{2023}), ISSN \bibinfo{issn}{2041-1723}, \urlprefix\url{https://doi.org/10.1038/s41467-023-42989-7}.

\bibitem[{\citenamefont{Yoshioka et~al.}(2024)\citenamefont{Yoshioka, Bernard, Wakamura, Hashisaka, Sasaki, Sasaki, Watanabe, Taniguchi, and Kumada}}]{Yoshioka2024}
\bibinfo{author}{\bibfnamefont{K.}~\bibnamefont{Yoshioka}}, \bibinfo{author}{\bibfnamefont{G.}~\bibnamefont{Bernard}}, \bibinfo{author}{\bibfnamefont{T.}~\bibnamefont{Wakamura}}, \bibinfo{author}{\bibfnamefont{M.}~\bibnamefont{Hashisaka}}, \bibinfo{author}{\bibfnamefont{K.-i.} \bibnamefont{Sasaki}}, \bibinfo{author}{\bibfnamefont{S.}~\bibnamefont{Sasaki}}, \bibinfo{author}{\bibfnamefont{K.}~\bibnamefont{Watanabe}}, \bibinfo{author}{\bibfnamefont{T.}~\bibnamefont{Taniguchi}}, \bibnamefont{and} \bibinfo{author}{\bibfnamefont{N.}~\bibnamefont{Kumada}}, \bibinfo{journal}{Nature Electronics} \textbf{\bibinfo{volume}{7}}, \bibinfo{pages}{537} (\bibinfo{year}{2024}), ISSN \bibinfo{issn}{2520-1131}, \urlprefix\url{https://doi.org/10.1038/s41928-024-01197-x}.

\bibitem[{\citenamefont{McIver et~al.}(2020)\citenamefont{McIver, Schulte, Stein, Matsuyama, Jotzu, Meier, and Cavalleri}}]{McIver2020}
\bibinfo{author}{\bibfnamefont{J.~W.} \bibnamefont{McIver}}, \bibinfo{author}{\bibfnamefont{B.}~\bibnamefont{Schulte}}, \bibinfo{author}{\bibfnamefont{F.-U.} \bibnamefont{Stein}}, \bibinfo{author}{\bibfnamefont{T.}~\bibnamefont{Matsuyama}}, \bibinfo{author}{\bibfnamefont{G.}~\bibnamefont{Jotzu}}, \bibinfo{author}{\bibfnamefont{G.}~\bibnamefont{Meier}}, \bibnamefont{and} \bibinfo{author}{\bibfnamefont{A.}~\bibnamefont{Cavalleri}}, \bibinfo{journal}{Nature Physics} \textbf{\bibinfo{volume}{16}}, \bibinfo{pages}{38} (\bibinfo{year}{2020}), ISSN \bibinfo{issn}{1745-2481}, \urlprefix\url{https://doi.org/10.1038/s41567-019-0698-y}.

\bibitem[{\citenamefont{Zhao et~al.}(2023)\citenamefont{Zhao, Wang, Chen, Zhang, Watanabe, Taniguchi, Zettl, and Wang}}]{Zhao2023}
\bibinfo{author}{\bibfnamefont{W.}~\bibnamefont{Zhao}}, \bibinfo{author}{\bibfnamefont{S.}~\bibnamefont{Wang}}, \bibinfo{author}{\bibfnamefont{S.}~\bibnamefont{Chen}}, \bibinfo{author}{\bibfnamefont{Z.}~\bibnamefont{Zhang}}, \bibinfo{author}{\bibfnamefont{K.}~\bibnamefont{Watanabe}}, \bibinfo{author}{\bibfnamefont{T.}~\bibnamefont{Taniguchi}}, \bibinfo{author}{\bibfnamefont{A.}~\bibnamefont{Zettl}}, \bibnamefont{and} \bibinfo{author}{\bibfnamefont{F.}~\bibnamefont{Wang}}, \bibinfo{journal}{Nature} \textbf{\bibinfo{volume}{614}}, \bibinfo{pages}{688} (\bibinfo{year}{2023}), ISSN \bibinfo{issn}{1476-4687}, \urlprefix\url{https://doi.org/10.1038/s41586-022-05619-8}.

\end{thebibliography}

\end{document}